\documentclass[preprint]{aastex}

\usepackage{amsmath}	
\usepackage{amssymb}	
\usepackage{graphicx,color}

\usepackage{natbib}

\newcommand{\vecB}{\boldsymbol{B}}
\newcommand{\ez}{\hat{\boldsymbol{e}}_z}

\newcommand{\rhoO}{\rho_0}
\newcommand{\rhoi}{\rho_i}
\newcommand{\rhoe}{\rho_e}

\newcommand{\vecxi}{\boldsymbol{\xi}}
\newcommand{\vecv}{\boldsymbol{v}}
\newcommand{\vecb}{\boldsymbol{b}}
\newcommand{\vecbperp}{\boldsymbol{b}_\perp}

\newcommand{\va}{v_{A}}
\newcommand{\vai}{v_{\rm Ai}}
\newcommand{\vae}{v_{\rm Ae}}

\newcommand{\kr}{k_r}
\newcommand{\ki}{k_i}
\newcommand{\ke}{k_e}

\newcommand{\kappae}{\kappa_e}

\newcommand{\Ci}{C_i}
\newcommand{\Ce}{C_e}

\newcommand{\cph}{c_{\rm ph}}
\newcommand{\cg}{c_g}

\newcommand{\tauAi}{\tau_{\rm Ai}}

\shorttitle{Dispersion of fast sausage wave trains in magnetic flux tubes}
\shortauthors{Oliver et al.}

\begin{document}

\title{PROPAGATION AND DISPERSION OF SAUSAGE WAVE TRAINS IN~MAGNETIC FLUX TUBES}

%

\author{R. Oliver}
\affil{Departament de F\'\i sica, Universitat de les Illes Balears, E-07122 Palma de Mallorca, Spain}
\email{e-mail: ramon.oliver@uib.es}

\author{M. S. Ruderman}
\affil{School of Mathematics and Statistics, University of Sheffield, Hicks Building, Hounsfield Road, Sheffield S3 7RH, UK \\ Space Research Institute (IKI), Russian Academy of Sciences, Moscow 117997, Russia}

\and
\author{J. Terradas}
\affil{Departament de F\'\i sica, Universitat de les Illes Balears, E-07122 Palma de Mallorca, Spain}

\begin{abstract}
{A localized perturbation of a magnetic flux tube produces a pair of wave trains that propagate in opposite directions along the tube. These wave packets disperse as they propagate, where the extent of dispersion depends on the physical properties of the magnetic structure, on the length of the initial excitation, and on its nature (e.g., transverse or axisymmetric). In \citet{oliver2014b} we considered a transverse initial perturbation, whereas the temporal evolution of an axisymmetric one is examined here. In both papers we use a method based on Fourier integrals to solve the initial value problem. Previous studies on wave propagation in magnetic wave guides have emphasized that the wave train dispersion is influenced by the particular dependence of the group velocity on the longitudinal wavenumber. Here we also find that long initial perturbations result in low amplitude wave packets and that large values of the magnetic tube to environment density ratio yield longer wave trains. To test the detectability of propagating transverse or axisymmetric wave packets in magnetic tubes of the solar atmosphere (e.g., coronal loops, spicules, or prominence threads) a forward modelling of the perturbations must be carried out. This is left for a future work.}
\end{abstract}


\section{INTRODUCTION}

\begin{figure}[ht!]
  \centerline{(a)\includegraphics[width=0.18\textwidth,angle=0]{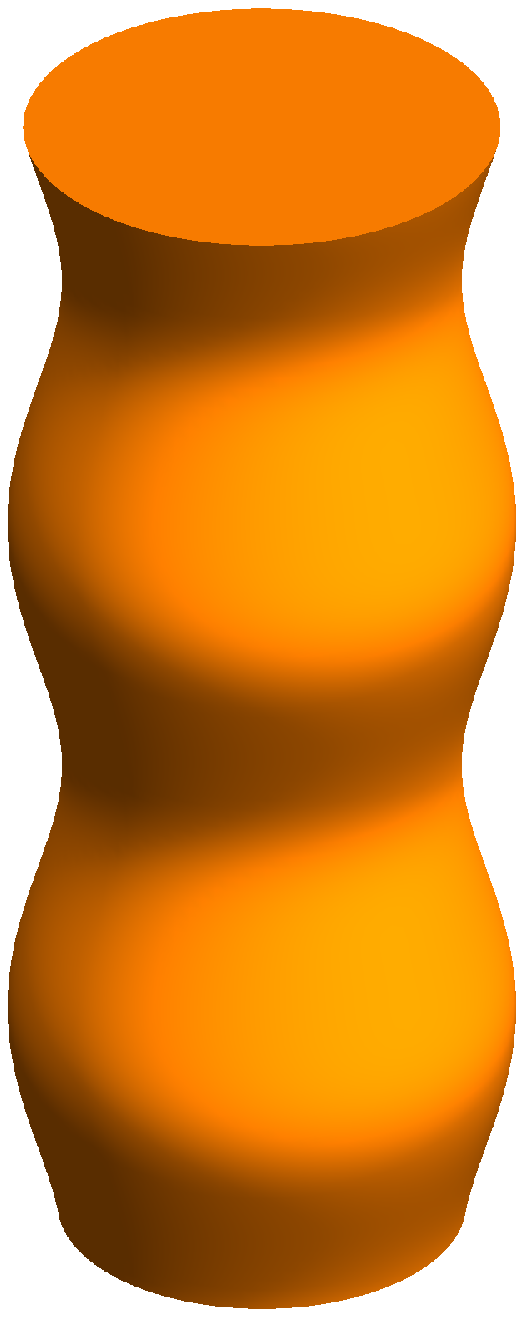}\hspace{1cm}(b)\includegraphics[width=0.18\textwidth,angle=0]{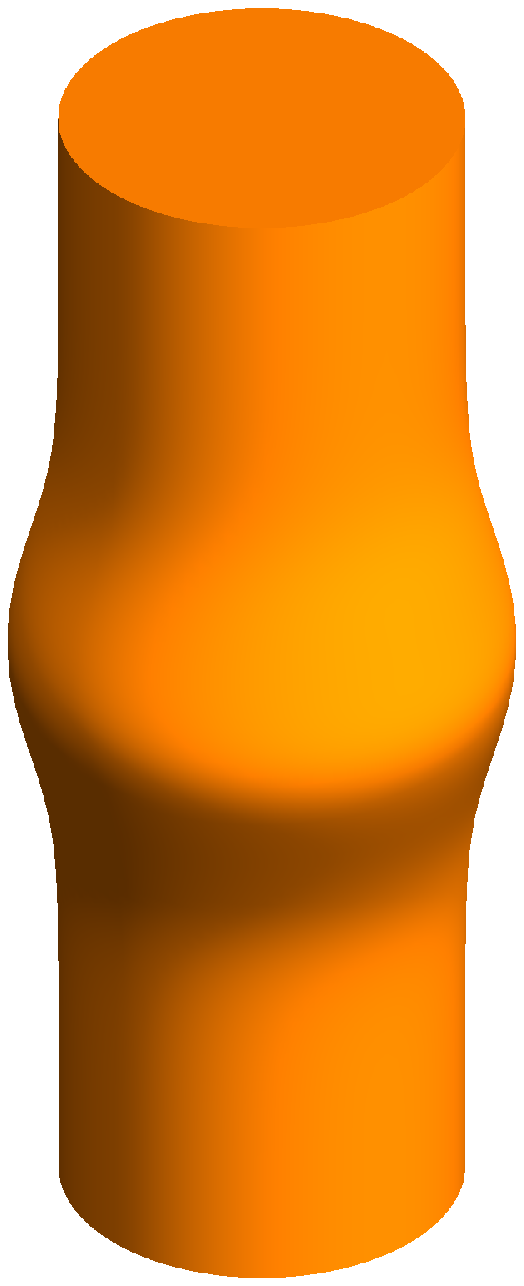}}
  \caption{Shape of the boundary of a uniform cylindrical magnetic flux tube subject to (a) a fast sausage normal mode perturbation and (b) a localized axisymmetric perturbation.}
  \label{fig_cylinder_sausage}
\end{figure}

The solar corona supports a rich variety of wave phenomena whose main features are determined by the magnetic field and plasma structuring. For our purposes it is important to distinguish between different wave manifestations such as driven and impulsively excited perturbations. Driven waves are often caused by a continuous, periodic exciter at the photospheric or chromospheric  level and propagate away from the place where they are generated. This is the case, for example, of waves propagating upwards along spicules \citep{okamoto2011,depontieu2012} or along the coronal magnetic field \citep{tomczyk2007,tomczyk2009}. The disturbances propagating in filament threads observed by \citet{lin2007,okamoto2007} are probably representative of driven waves too, although the exciting agent is not clear in this case. Rapidly propagating quasi-periodic coronal disturbances \citep{liu2011,liu2012b,shenliu2012,yuan2013,shen2013,nistico2014} also seem to be caused by a continuous driver that in this case is associated to a flare or the initiation of a CME. For a modelling of these events see \citet{ofman2011,nistico2014}. When a magnetic flux tube (e.g., spicule, coronal loop, or prominence thread) is subject to a periodic driving it shows a distinct response. An axisymmetric driver that periodically compresses and expands the tube at one position leads to the propagation of a sausage-type perturbation (see Figure~\ref{fig_cylinder_sausage}(a)). A periodic transverse shaking of the tube results in a kink-type perturbation by which the tube boundary acquires the shape of a snake \citep[see Figure~1(a) of][ hereafter Paper~I]{oliver2014b}. A torsional driver leads to Alfv\'en wave propagation along the tube with no apparent modification of its boundary.

An impulsive energy release can lead to both standing and propagating disturbances. The transverse coronal loop oscillations studied by \citet{aschwanden1999,nakariakov1999} constitute a good example of the first type of phenomenon, for which there are many observational studies in the literature. For recent observations using the Atmospheric Imaging Assembly (AIA) on board the {\it Solar Dynamics Observatory}, see \citet{nistico2013,anfinogentov2013}. Our interest in this work is in the second type of event, i.e., a magnetic flux tube perturbed by a sudden, localized expansion or compression such as that of Figure~\ref{fig_cylinder_sausage}(b) (see also Figure~1(b) of Paper~I for a localized transverse excitation). In a dispersive medium, such initial pulse disperses and transforms into a wave train that induces quasi-periodic plasma oscillations with a range of periods (rather than the single-period oscillations of driven propagating waves or standing oscillations). Hence, the time signal (e.g., Doppler shift or spectral line intensity) collected at a given position shows a characteristic ``tadpole'' shape in its wavelet diagram \citep[e.g.,][]{nakariakov2004,jelinek2010}. The important difference between driven and impulsively excited waves, then, is that the first ones have the period of the driving agent, whereas the second ones display a continuous variation of the period that depends on the size of the initial perturbation and the physical properties of the medium.

Several observations of quasi-periodic oscillations have been ascribed to the dispersive behavior of propagating wave packets. \citet{williams2001,williams2002,katsiyannis2003} used the Solar Eclipse Corona Imaging System (SECIS) to record white light and coronal green line images of the solar corona during a solar eclipse. They studied the intensity variations in the vicinity of an active region and detected periodic  variations with period in the range 4--7~s that propagate through the loop apex with a speed of 2100~km~s$^{-1}$. The wavelet spectra of signals at different points along the wave path display the ``tadpole'' shape mentioned above. This wavelet feature is also present in solar decimetric radio bursts \citep{meszarosova2009,meszarosova2011,karlicky2013} and in radio sources lying in a magnetic fan structure \citep{meszarosova2013}.

The impulsive excitation of fast magnetohydrodynamic waves, and their subsequent evolution as an oscillatory wave train, has been addressed for different magnetic structures: fast sausage wave trains in a coronal slab  \citep{murawski1993b, murawski1993d,nakariakov2004,nakariakov2005, jelinek2010,nakariakov2012, meszarosova2014}, in a current sheet \citep{jelinek2012a,jelinek2012b, karlicky2013,meszarosova2014}, and in a magnetic funnel \citep{pascoe2013}. Impulsively generated fast sausage and kink wave packets in coronal holes have also been studied by \citet{pascoe2014}. Our paper is concerned with the temporal evolution of a concentrated axisymmetric compression or expansion of a uniform, cylindrical magnetic tube; see Figure~\ref{fig_cylinder_sausage}(b). This initial disturbance leads to the propagation of a fast sausage wave train along the tube. The propagation and dispersion of wave packets in magnetic slabs has already been investigated numerically in slab geometry, but not in cylindrical geometry.

This is the outline of our paper. Section~\ref{sect_gov_eqns} contains the governing equations of linear fast perturbations in a plasma cylinder in the $\beta=0$ limit. In Sections~\ref{sect_proper} and \ref{sect_improper} we give a description of fast sausage proper and improper modes, whose spectra are discrete and continuous, respectively. The initial perturbation is used in Section~\ref{sect_fourier} to obtain analytical expressions for the time evolution of the perturbed variables and the results are presented in Section~\ref{sect_results}.


\section{EQUILIBRIUM AND GOVERNING EQUATIONS FOR LINEAR PERTURBATIONS IN A COLD PLASMA CYLINDER}\label{sect_gov_eqns}

We consider a homogeneous cylindrical magnetic tube embedded in a uniform environment. In this work we use cylindrical coordinates $(r,\varphi,z)$, with the $z$-axis pointing along the cylinder axis. The magnetic field is uniform in the whole space and parallel to the $z$-axis, and so can be written as $\vecB=B\ez$. The equilibrium density is

\begin{equation}\label{rho0}
\rhoO(r) = 
   \begin{cases}
    \rhoi, & r \leq a, \vspace{1mm} \\
    \rhoe, & r > a,
  \end{cases}
\end{equation}

\noindent
where the subscripts `$i$' and `$e$' denote internal and external values, respectively, and $a$ is the tube radius.

Next we consider small perturbations about this equilibrium, governed by the linear, ideal magnetohydrodynamic equations, in the cold plasma approximation \citep[Equations~(1)--(3) of ][]{ruderman2009}. In our work $\vecv$ is the velocity perturbation, $\vecxi$ the Lagrangian plasma displacement, $\vecb$ the magnetic field perturbation, $P=Bb_z/\mu_0$ the magnetic pressure perturbation (with $\mu_0$ the magnetic permeability of free space), and $\rho$ the density perturbation. Using Equations~(1)--(3) of \citet{ruderman2009} it is straightforward to see that, if $\xi_z$ and $v_z$ are initially zero, then they vanish for all times. In addition, perturbations are governed by

\begin{equation}\label{conti_eqn}
\rho=\frac{1}{\va^2}P,
\end{equation}

\begin{equation}\label{mom_eqn}
\frac{\partial^2\vecxi}{\partial t^2}=-\frac{1}{\rhoO}\nabla_\perp P+\va^2\frac{\partial^2\vecxi}{\partial z^2},
\end{equation}

\begin{equation}\label{b_eqn}
\vecbperp=B\frac{\partial\vecxi}{\partial z},
\end{equation}

\begin{equation}\label{P_eqn}
P=-\rhoO\va^2\nabla\cdot\vecxi,
\end{equation}

\begin{equation}\label{v_eqn}
\vecv=\frac{\partial\vecxi}{\partial t},
\end{equation}

\noindent
where we have written $\vecbperp=\vecb-b_z\ez$ and

\[
\nabla_\perp=\nabla-\ez\frac{\partial}{\partial z}.
\]

\noindent
Moreover, $\va$ is the Alfv\'en speed, whose values inside and outside the magnetic cylinder are respectively given by

\begin{equation}\label{eq_alfven}
{\vai^2} =\frac{B^2}{\mu_0\rhoi}, \qquad \vae^2 = \frac{B^2} {\mu_0\rhoe}.
\end{equation}

\noindent
In this paper we consider a magnetic tube that is denser than the environment, i.e., $\rhoi>\rhoe$ and $\vai<\vae$.

\section{PROPER (OR DISCRETE) FAST SAUSAGE MODES}\label{sect_proper}

\subsection{Eigenfunctions}\label{sect_proper_eigen}

For sausage perturbations, $P$, $\rho$, $\vecxi$, $\vecb$, and $\vecv$ have no azimuthal dependence and as a consequence the $\varphi$-components of Equations~(\ref{mom_eqn}), (\ref{b_eqn}), and (\ref{v_eqn}) dictate that $\xi_\varphi$, $b_\varphi$, and $v_\varphi$ vanish (provided that they are initially zero). In addition, the variables are of the form

\begin{equation}\label{rzt_dependence}
P(t,r,\varphi,z) = \hat{P}(r)\exp[i(-\omega t + k z)],
\end{equation}

\noindent
and similarly for $\rho$, $\xi_r$, $b_r$, and $v_r$. The eigenfunctions of fast magnetohydrodynamic (MHD) modes in a magnetic cylinder have the azimuthal dependence $\exp(im\varphi)$, with $m=0$ for sausage modes.

Equation~(\ref{conti_eqn}) gives the radial dependence of the density perturbation

\begin{equation}\label{rho_pert}
\hat{\rho} = \frac{1}{\va^2}\hat{P}.
\end{equation}

\noindent
Equation~(\ref{mom_eqn}) leads to

\begin{equation}\label{xi_pert}
\hat{\xi}_r = \frac1{\rho_0 \va^2 k_r^2}\frac{d\hat{P}}{d r},
\end{equation}

\noindent
where the radial wavenumber, $\kr$, is given by

\begin{equation}\label{kr}
\kr^2 = \frac{\omega^2-k^2\va^2}{\va^2}.
\end{equation}

\noindent
Equations~(\ref{b_eqn}) and (\ref{v_eqn}) give the radial components of the magnetic field and velocity perturbations in terms of the radial plasma displacement,

\begin{equation}\label{b_pert}
\hat{b}_r = ikB\hat{\xi}_r,
\end{equation}

\begin{equation}\label{v_pert}
\hat{v}_r = -i\omega\hat{\xi}_r.
\end{equation}

\noindent
Finally, from Equations~(\ref{P_eqn}) and (\ref{xi_pert}) we obtain

\begin{equation}\label{equ_for_hatP}
\frac{d^2\hat{P}}{d r^2} + \frac1r\frac{d\hat{P}}{d r} + k_r^2 \hat{P} = 0.
\end{equation}

To determine the solution to Equation~(\ref{equ_for_hatP}) we need to take into account the possible signs of~$\kr^2$ inside and outside the magnetic cylinder (the internal and external radial wavenumbers are here denoted as $\ki$ and $\ke$, respectively). For proper fast sausage modes $k\vai < \omega < k \vae$ and this implies that $\ki^2>0$, $\ke^2<0$. Then, the solution to Equation~(\ref{equ_for_hatP}) regular at $r=0$ and that does not diverge as $r\to\infty$ is

\begin{equation}\label{hatP}
\hat{P}(r) = 
   \begin{cases}
    \Ci J_0(\ki r), & r \leq a, \vspace{1mm} \\
    \Ce K_0(\kappae r), & r \geq a,
  \end{cases}
\end{equation}

\noindent
where $\Ci$ and $\Ce$ are two constants, $J_m$ is the Bessel function of the first kind and order $m$, and $K_m$ is the modified Bessel function of the second kind and order $m$. Moreover,

\begin{equation}\label{ki}
\ki^2 = \frac{\omega^2-k^2\vai^2}{\vai^2},
\end{equation}

\noindent
and $\kappae^2=-\ke^2$ is

\begin{equation}\label{kappae}
\kappae^2 = -\frac{\omega^2-k^2\vae^2}{\vae^2}.
\end{equation}

\noindent
It is important to mention that when using these equations to determine the radial wavenumbers, the positive value is taken, i.e., $\ki>0$, $\kappae>0$. An algebraic expression for $\Ci$ and $\Ce$ is obtained by imposing the continuity of $\hat{P}(r)$ at $r=a$. The remaining constant can be imposed arbitrarily. Here we use expressions analogous to those of Paper~I, that is

\begin{equation}\label{CiCe}
\Ci=\rhoi \vai^2 K_0(\kappae a), \qquad
\Ce=\rhoe \vae^2 J_0(\ki a).
\end{equation}

Now that the function $\hat{P}(r)$ is determined, all other perturbations can be obtained from Equations~(\ref{rho_pert}), (\ref{xi_pert}), (\ref{b_pert}), and (\ref{v_pert}). We write the expression of the radial plasma displacement because it will be used later,

\begin{equation}\label{xir_of_r_primes}
\hat{\xi}_r(r) = 
   \begin{cases}
    \hphantom{-}\ki^{-1} K_0(\kappae a) J'_0(\ki r), & r \leq a, \vspace{1mm}\\
    -\kappae^{-1}  J_0(\ki a) K'_0(\kappae r), & r \geq a,
  \end{cases}
\end{equation}

\noindent where a prime indicates a derivative. From now on we use the relations \citep{abramowitz1964}

\begin{equation}\label{J0pK0p}
J_0'(x)=-J_1(x), \qquad K_0'(x)=-K_1(x),
\end{equation}

\noindent and write Equation~(\ref{xir_of_r_primes}) as

\begin{equation}\label{xir_of_r}
\hat{\xi}_r(r) = 
   \begin{cases}
    -\ki^{-1} K_0(\kappae a) J_1(\ki r), & r \leq a, \vspace{1mm}\\
    \hphantom{-}\kappae^{-1}  J_0(\ki a) K_1(\kappae r), & r \geq a.
  \end{cases}
\end{equation}

\subsection{Dispersion Relation, Phase Speed, and Group Velocity}

Imposing the continuity of the radial displacement at $r=a$ leads to the dispersion relation for proper fast sausage modes,

\begin{equation}\label{dr_kisq_pos}
\frac{J_1(\ki a)}{\ki J_0(\ki a)} + 
   \frac{K_1(\kappae a)}{\kappae K_0(\kappae a)} = 0.
\end{equation}

\noindent
In this work the dispersion relation is used to obtain the frequency of a proper mode with a given longitudinal wavenumber, $k$. Therefore, we often write the frequency as $\omega(k)$.

Moreover, the phase speed and group velocity are defined in the usual manner,

\begin{equation}\label{cph_cg}
\cph(k) = \frac\omega k, \qquad \cg(k)=\frac{\partial\omega}{\partial k}.
\end{equation}

\noindent
Although the phase speed is not used in our calculations nor in the interpretation of results, it is also given here because it is an important parameter of proper modes. The main features of $\omega$, $\cph$, and $\cg$ of proper sausage modes in a magnetic cylinder have been discussed in previous works \citep[e.g.,][]{edwinroberts1983,roberts1984,cally1986}. They are summarised below.

First, we discuss some properties of the eigensolutions for the possible sign combinations of $k$ and $\omega$; they will be useful when the method of Fourier integrals is applied to the wave propagation problem. It is clear that the two pairs of values $(k,\omega)$ and $(-k,-\omega)$ correspond to the same eigenmode, propagating in a given direction along the $z$-axis. It is obvious that the solutions presented above have this property: if the signs of both $k$ and $\omega$ are simultaneously changed, then Equations~(\ref{ki}) and (\ref{kappae}) show that the internal and external radial wavenumbers are not modified. As a result, these new values of the longitudinal wavenumber and frequency also satisfy the dispersion relation. In other words, if $\omega(k)$ is the frequency of a proper mode, then $-\omega(-k)$ is also a frequency of a proper mode. Using Equations~(\ref{rho_pert}), (\ref{b_pert}), (\ref{v_pert}), (\ref{hatP}), and (\ref{xir_of_r}), we see that the perturbations of both modes are identical. Finally, we also have $\cph(k)=\cph(-k)$ and $\cg(k)=\cg(-k)$. Next let us consider a fixed $k$: if $\omega(k)$ is a solution to the dispersion relation, then $-\omega(k)$ corresponds to the eigenmode propagating in the opposite direction along the magnetic tube. Again, the two eigenmodes have the same eigenfunctions, but opposite phase speed and group velocity. Thus, in the rest of this section it is enough to concentrate in $k\geq 0$, $\omega> 0$. Both positive and negative values of $k$ and $\omega$ will be taken into account when deriving the solution to the initial value problem by the method of Fourier integrals (Section~\ref{sect_fourier}).

The magnetic cylinder supports an infinite number of fast sausage radial harmonics, characterised by the number of nodes of $\hat P(r)$. All radial harmonics have a node at the tube axis, $r = 0$. This is the only node for the fundamental sausage eigenmode. The first overtone has two nodes in the radial direction, etc., and all these nodes are inside the magnetic cylinder. An additional feature of proper fast sausage modes is that they only exist for longitudinal wavenumbers exceeding the cut-off value \citep{roberts1984,nakariakov_verwichte2005},

\begin{equation}\label{eq_kcj}
k_{cj} = \frac{j_{0j}\vai}{a\sqrt{\vae^2 - \vai^2}} ,
\end{equation}

\noindent
where $j_{0j}$ is the $j$\/th zero (in increasing order) of $J_0(x)$. Here $j=1,2,\ldots$ correspond to the fundamental fast sausage mode, its first radial overtone, etc.

\begin{figure}[ht!]
    \scriptsize{(a)}\hspace{-10ex}
    \includegraphics[width=0.35\textwidth,angle=-90]{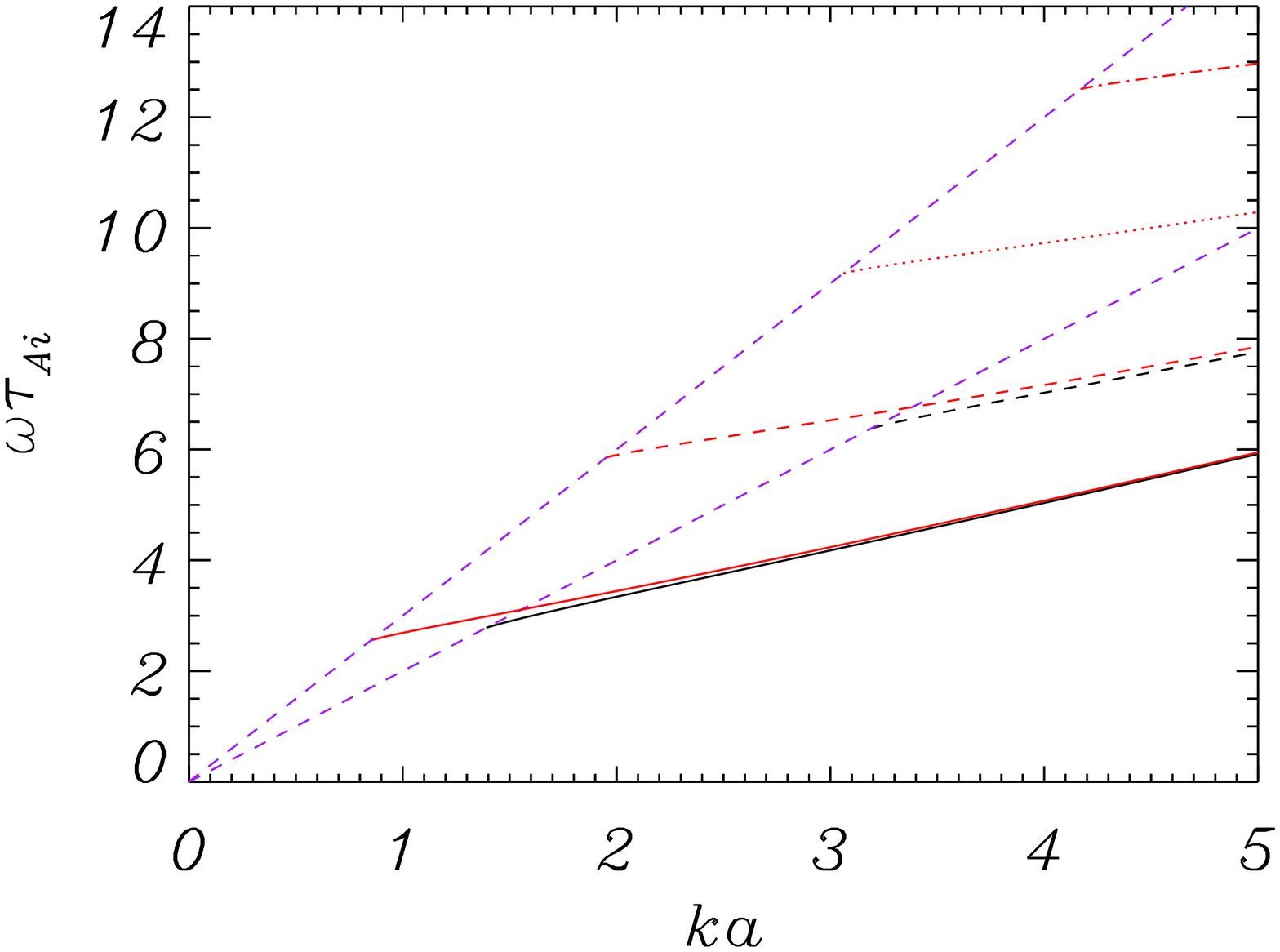} \\
    \scriptsize{(b)}\hspace{-10ex}
    \includegraphics[width=0.35\textwidth,angle=-90]{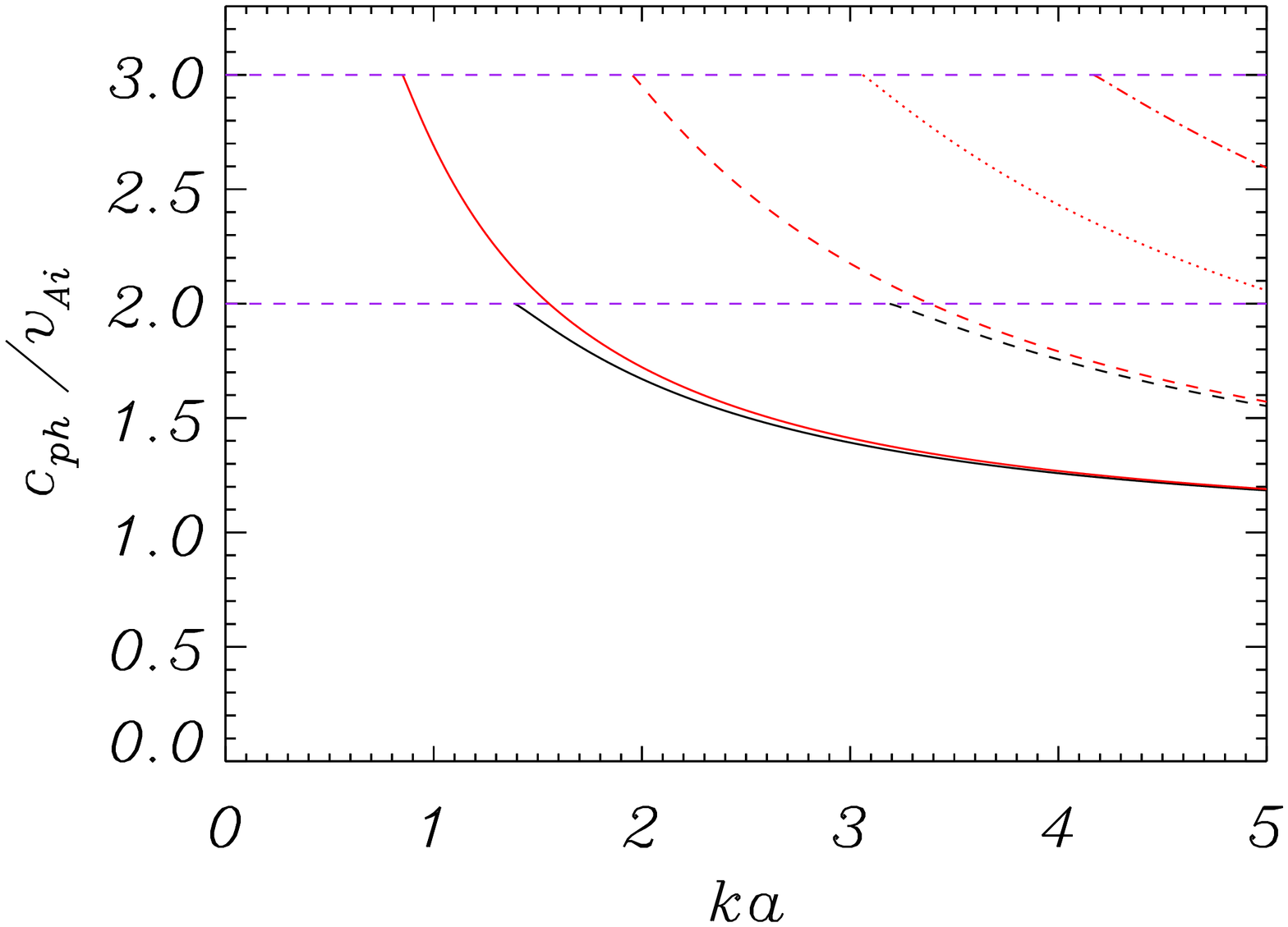} \\
    \scriptsize{(c)}\hspace{-10ex}
    \includegraphics[width=0.35\textwidth,angle=-90]{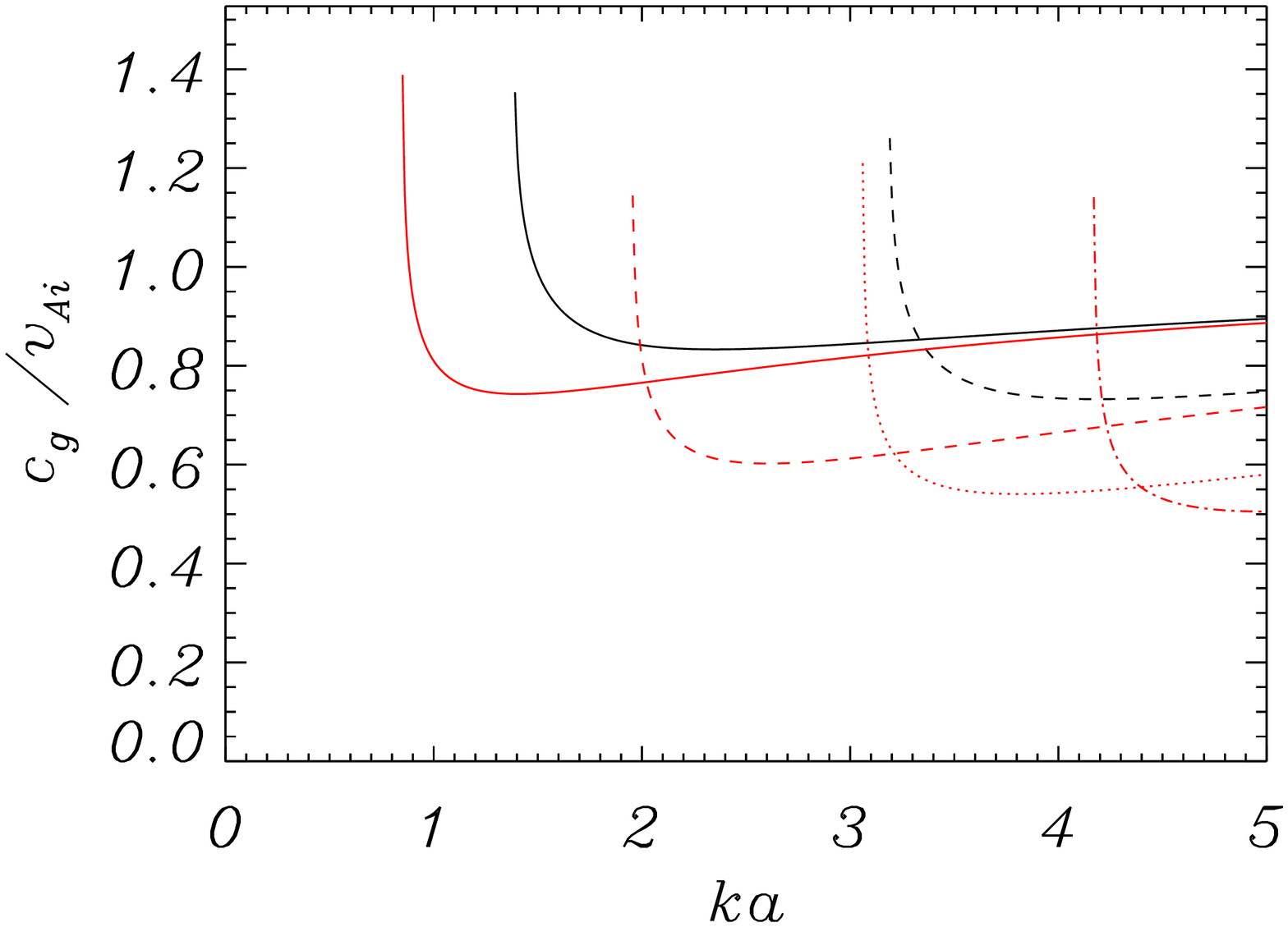}
  \caption{(a) Frequency, (b) phase speed, and (c) group velocity vs. the longitudinal wavenumber for the fundamental fast sausage mode (solid line) and its first overtones in the radial direction (dashed, dotted, and dash-dotted lines). Black and red colors respectively correspond to the equilibrium density ratio $\rhoi/\rhoe=4$ and 9, i.e., to the Alfv\'en speed ratio $\vae/\vai=2$ and~3. The purple lines in (a) and (b) correspond to the frequency $\omega=k\vae$. To obtain the dimensionless quantities, the magnetic cylinder radius ($a$), Alfv\'en speed ($\vai$), and transit time ($\tauAi=a/\vai$) have been used.}
  \label{fig_dr_over}
\end{figure}

In Figure~\ref{fig_dr_over} the frequency, phase speed, and group velocity of the fundamental fast sausage eigenmode and its first overtones are shown for two values of the density ratio, $\rhoi/\rhoe$. The dispersion diagram (Figure~\ref{fig_dr_over}(a)) is rarely found in the literature. It shows that all proper modes, including the fundamental one, have a wavenumber cut-off and that their frequency satisfies $k\vai<\omega<k\vae$. One can see that increasing the density ratio raises the upper frequency limit $k\vae$ and so more overtones are present in this diagram. It is worth to mention that linear waves with $k\vae\leq\omega$ also exist, but instead of discrete values of $\omega$ these solutions can have any frequency (see Section~\ref{sect_improper}). Figure~\ref{fig_dr_over}(b) has been presented many times in other works and so is not discussed here. It is enough to note that the phase speed lies between the internal and external Alfv\'en speeds. Finally, Figure~\ref{fig_dr_over}(c) shows the group velocity as a function of $k$. When the wavenumber approaches its cut-off value from the right, $\cg$ rapidly increases and tends to $\vae$, i.e., it tends to 3 and 2 for the red and black lines, respectively. For larger wavenumbers, the group velocity decreases to its minimum and finally increases slowly towards $\vai$. In Appendix~\ref{app_large_k} we show that the phase speed tends to $\vai$ from above as $ka\rightarrow\infty$. Moreover, the group velocity tends to $\vai$ from below for very large wavenumbers. This behavior of $\cph$ and $\cg$ is in agreement with Figures~\ref{fig_dr_over}(b) and (c).

\section{IMPROPER (OR CONTINUOUS) FAST SAUSAGE MODES}\label{sect_improper}

When a fixed $k$ is considered, proper modes have discrete frequencies in the range $(-k\vae,k\vae)$ (see purple dashed lines of Figure~\ref{fig_dr_over}(a)). Improper mode eigenfunctions can also be expressed in the form of Equation~(\ref{rzt_dependence}), but they can have any frequency in the ranges $(-\infty,-k\vae]$ and $[k\vae,\infty)$. Paper~I contains a detailed discussion of kink improper modes properties. All the arguments presented in Paper~I are valid for improper sausage modes and so are not repeated here. The total pressure and radial displacement perturbations are given by

\begin{equation}\label{hatP_im}
\hat{P}(r) = 
   \begin{cases}
    \rhoi \vai^2 a^2 \ke^2 J_0(\ki r), & r \leq a, \vspace{1mm} \\
    \rhoe \vae^2 a^2 \ke^2 [C_J J_0(\ke r) + C_Y Y_0(\ke r)], & r \geq a,
  \end{cases}
\end{equation}

\begin{equation}\label{xir_of_r_im}
\hat{\xi}_r(r) = 
   \begin{cases}
    \hphantom{xxxx}-a^2 \ke^2\ki^{-1} J_1(\ki r), & r < a, \vspace{1mm}\\
    -a^2\ke[C_J J_1(\ke r) + C_Y Y_1(\ke r)], & r > a,
  \end{cases}
\end{equation}

\noindent
where $Y_m$ is the Bessel function of the second kind and order $m$, and $C_J$ and $C_Y$ are two constants,

\begin{equation}\label{C_J_C_Y}
  \begin{array}{l}\displaystyle
    C_J = \frac{\pi a\ke}{2\ki}\left[-\ki J_0(\ki a) Y_1(\ke a) + \ke J_1(\ki a) Y_0(\ke a)\right],
       \vspace{2mm}\\ \displaystyle
    C_Y = \frac{\pi a\ke}{2\ki}\left[-\ke J_1(\ki a) J_0(\ke a) + \ki J_0(\ki a) J_1(\ke a)\right] . 
  \end{array}
\end{equation} 

\noindent
All other perturbations can be derived from the expressions of Section~\ref{sect_proper_eigen}.

\section{SOLUTION BY FOURIER INTEGRALS}\label{sect_fourier}

\subsection{Solution to Initial Value Problem}\label{sect_amplitudes}

In this section we derive the expressions governing the propagation of an axisymmetric impulsive excitation such as that of Figure~\ref{fig_cylinder_sausage}(a). The procedure to obtain the evolution of the perturbations after the magnetic tube suffers a localized transverse displacement is described in detail in Paper~I, where all the relevant information can be found; here only the main equations are given. The initial value problem is specified by two initial conditions on the radial displacement,

\begin{equation}\label{init_cond}
\begin{array}{l}
\xi_r(t=0,r,z) = f(r,z), \\ [2ex]
\dfrac{\partial\xi_r}{\partial t} (t=0,r,z)= g(r,z).
\end{array}
\end{equation}

To impose these initial conditions we write the Fourier transform of $\xi_r(t,r,z)$ with respect to~$z$,

\begin{equation}\label{Four_trans-main}
\tilde{\xi}_r(t,r,k) = \int_{-\infty}^\infty
  \xi_r(t,r,z) e^{-ikz}\,dz.
\end{equation}

\noindent
For a given longitudinal wavenumber, $\tilde{\xi}_r(t,r,k)$ can be expressed as a linear combination of proper and improper eigenfunctions,

\begin{align}\label{expand_eigen-main}
\tilde{\xi}_r(t,r,k) &= \sum_{j=1}^N \left[A_j^+(k)e^{-i\omega_j(k)t} + 
   A_j^-(k)e^{i\omega_j(k)t}\right]\hat{\xi}_j(r,k) \nonumber \\
&+ \int_{|k|\vae}^\infty\left[A_\omega^+(k)e^{-i\omega t} + 
   A_\omega^-(k)e^{i\omega t}\right]\hat{\xi}_\omega(r,k)\,d\omega ,
\end{align}

\noindent
where $\omega_j(k)>0$ and $\omega>0$ are assumed with no loss of generality. In this expression, the fundamental fast sausage mode ($j=1$) and its radial overtones ($j=2,3,\ldots$) have frequencies $\pm\omega_j(k)$, which are the solutions to the dispersion relation (Equation~(\ref{dr_kisq_pos})), and their radial displacement, $\hat{\xi}_j(r,k)$, is given by Equation~(\ref{xir_of_r}). The contribution of each of these proper modes to Equation~(\ref{expand_eigen-main}) depends on the amplitude $A_j^\pm(k)$, where the $+$ and $-$ signs correspond to propagation in opposite directions along the magnetic tube. Moreover, $N$ is the number of proper sausage modes for the selected longitudinal wavenumber. For $|k|\leq k_{c1}$ no proper eigenmodes exist and so $N=0$, which means that the sum is not present in Equation~(\ref{expand_eigen-main}); for $k_{c1}<|k|\leq k_{c2}$ only the fundamental fast sausage mode exists and then $N=1$; and so on. Regarding improper modes, their contribution to Equation~(\ref{expand_eigen-main}) has the form of an integral because they have a continuous variation of frequency for a given $k$. The improper modes eigenfunction $\hat{\xi}_\omega(r,k)$ has been presented in Equation~(\ref{xir_of_r_im}).

We next define the Hilbert space of functions on the interval $[0, \infty)$ by the scalar product

\begin{equation}\label{scalar_prod}
\langle\eta,\zeta\rangle = \int_0^\infty \rho_0(r) (\eta\zeta^*) r\,dr .
\end{equation}

\noindent
With this definition, the eigenfunctions satisfy the orthogonality conditions

\begin{equation}\label{scalar}
\begin{array}{l}
\langle\hat{\xi}_j,\hat{\xi}_l\rangle = 0 \quad (j \neq l) ,
    \vspace*{2mm}\\
\langle\hat{\xi}_j,\hat{\xi}_\omega\rangle = 0 , \vspace*{2mm}\\
\langle\hat{\xi}_\omega,\hat{\xi}_{\omega'}\rangle = 
   q(\omega)\delta(\omega-\omega'), 
\end{array} 
\end{equation}

\noindent
where $\omega,\omega'>0$ and

\begin{equation}\label{q_omega-main}
q(\omega) = \frac{\rho_e a^4\vae^2\ke^2}{\omega}\left(C_J^2 + C_Y^2\right) ,
\end{equation}

\noindent see Appendix~\ref{app_q_of_omega}.

To determine the functions $A_j^\pm(k)$ ($j = 1,\dots, N$) and $A_\omega^\pm(k)$ we use the initial conditions. Taking $t = 0$ in Equation~(\ref{expand_eigen-main}) we obtain 

\begin{align}\label{init_expand_xi}
\tilde{f}(r,k) &=\sum_{j=1}^N \left[A_j^+(k) + 
   A_j^-(k)\right]\hat{\xi}_j(r,k) \nonumber \\
&+ \int_{|k|\vae}^\infty\left[A_\omega^+(k) + 
   A_\omega^-(k)\right]\hat{\xi}_\omega(r,k)\,d\omega ,  
\end{align}

\noindent
$\tilde{f}(r,k)$ being the Fourier transform with respect to $z$ of $f(r,z)$. Now, differentiating Equation~(\ref{expand_eigen-main}) with respect to $t$ and substituting $t=0$ yields

\begin{align}\label{init_expand_der}
\tilde{g}(r,k) &=- i\sum_{j=1}^N \omega_j(k)\left[A_j^+(k) - 
   A_j^-(k)\right]\hat{\xi}_j(r,k) \nonumber \\
&- i\int_{|k|\vae}^\infty \omega\left[A_\omega^+(k) - 
   A_\omega^-(k)\right]\hat{\xi}_\omega(r,k)\,d\omega ,  
\end{align}

\noindent
with $\tilde{g}(r,k)$ the Fourier transform of $g(r,z)$ with respect to $z$\/.

The next step is to take the scalar product of Equations~(\ref{init_expand_xi}) and (\ref{init_expand_der}) with $\hat{\xi}_j(r,k)$. Using Equation~(\ref{scalar}), this results in two algebraic expressions for $A_j^\pm(k)$ that lead to

\begin{equation}\label{A_pm-main}
A_j^\pm(k) = \frac{\mathcal{N}^\pm_j(k)}{2\omega_j(k)\mathcal{D}_j(k)} , \qquad (j = 1,2,\dots,N),
\end{equation}

\noindent
with

\begin{align}\label{A_numer-main}
\mathcal{N}^\pm_j(k) &=
   -\frac{\rhoi K_0(\kappae a)}{\ki^2} \int_0^a \left[(\omega_j \tilde{f} \pm 
   i\tilde{g}) \ki r J_1(\ki r) \right]\,dr \nonumber \\
&+ \frac{\rhoe J_0(\ki a)}{\kappae^2} \int_a^\infty \left[(\omega_j \tilde{f}\pm
   i\tilde{g}) \kappae r K_1 (\kappae r) \right]\,dr ,
\end{align}

\begin{align}\label{A_denom-main_int}
\mathcal{D}_j(k) &= \frac{\rhoi K_0^2(\kappae a)}{\ki^4} 
   \int_0^a (\ki r)^2 J_1^2(\ki r) \frac{dr}r \nonumber \\
&+ \frac{\rhoe J_0^2(\ki a)}{\kappae^4}
   \int_a^\infty (\kappae r)^2 K_1^2(\kappae r) \frac{dr}r  .
\end{align}

\noindent Analytical expressions for the integrals in this formula can be obtained. Using Equation~(\ref{J0pK0p}) and the relations \citep{abramowitz1964}

\begin{equation}\label{derivJ1K1}
\begin{array}{l}
xJ_1'(x)=xJ_0(x)-J_1(x), \\ [2ex]
xK_1'(x)=-xK_0(x)-K_1(x),
\end{array}
\end{equation}

\noindent we obtain

\begin{equation}
\begin{array}{l}
\dfrac{d}{dx}\left\{x^2\left[J_0^2(x)+J_1^2(x)\right]\right\}=2xJ_0^2(x), \\ [2ex]
\dfrac{d}{dx}\left\{x^2\left[K_0^2(x)-K_1^2(x)\right]\right\}=2xK_0^2(x).
\end{array}
\end{equation}

\noindent With the aid of these formulae we get

\begin{align}
\int_0^a (\ki r)^2 J_1^2(\ki r) \frac{dr}r &=\frac{\ki^2a^2}{2}\left[J_0^2(\ki a)+J_1^2(\ki a)\right] \nonumber \\
&-\ki a J_0(\ki a) J_1(\ki a),
\end{align}

\begin{align}
\int_a^\infty (\kappae r)^2 K_1^2(\kappae r) \frac{dr}r &=\frac{\kappae^2a^2}{2}\left[K_0^2(\kappae a)-K_1^2(\kappae a)\right] \nonumber \\
&+\kappae a K_0(\kappae a) K_1(\kappae a).
\end{align}

\noindent Inserting these expressions in Equation~(\ref{A_denom-main_int}) $\mathcal{D}_j(k)$ can be cast as

\begin{align}\label{A_denom-main}
\mathcal{D}_j(k) &= \frac{\rhoi K_0^2(\kappae a)}{\ki^4} 
   \bigg\{\frac{\ki^2a^2}{2}\left[J_0^2(\ki a)+J_1^2(\ki a)\right] \nonumber \\
   & \hspace{15ex} -\ki a J_0(\ki a) J_1(\ki a)\bigg\} \nonumber \\
&+ \frac{\rhoe J_0^2(\ki a)}{\kappae^4}
   \bigg\{\frac{\kappae^2a^2}{2}\left[K_0^2(\kappae a)-K_1^2(\kappae a)\right] \nonumber \\
& \hspace{15ex} +\kappae a K_0(\kappae a) K_1(\kappae a)\bigg\}.
\end{align}

\noindent Furthermore, the amplitudes of improper modes, $A_\omega^\pm(k)$, are obtained in a similar fashion by taking the scalar product of Equations~(\ref{init_expand_xi}) and (\ref{init_expand_der}) with $\hat{\xi}_\omega(r,k)$. Then we have

\begin{equation}\label{A_pm_om-main}
A_\omega^\pm(k) = \frac{\mathcal{N}^\pm_\omega(k)}{2\omega q(\omega)} ,
\end{equation}

\noindent
where

\begin{align}\label{A_numer_om-main}
\mathcal{N}^\pm_\omega(k) &=
   -\frac{\rhoi a^2\ke^2}{\ki^2} \int_0^a (\omega \tilde{f} \pm 
   i\tilde{g}) \ki r J_1(\ki r) \,dr  \nonumber \\
&- \rhoe a^2 \int_a^\infty (\omega\tilde{f} \pm i\tilde{g}) \ke r 
    \left[C_J J_1(\ke r) \right. \nonumber \\
& \hspace{25ex} \left. +\, C_Y Y_1(\ke r)\right] \,dr .
\end{align}

We can finally present the main expression of this paper. The inverse Fourier transform of $\tilde{\xi}_r(t,r,k)$ with respect to $k$ yields the radial displacement as a function of $r$ and $z$,

\begin{equation}\label{Four_trans_inv-main}
\xi_r(t,r,z) = \frac1{2\pi}\int_{-\infty}^\infty
   \tilde{\xi}_r(t,r,k)e^{ikz}\,dk,
\end{equation}

\noindent where $\tilde{\xi}_r(t,r,k)$ is given in Equation~(\ref{expand_eigen-main}). All other perturbations (i.e., magnetic field, velocity, and density) are described by analogous formulas.

\subsection{Amplitudes for Specific Initial Conditions}

The expressions presented in Section~\ref{sect_amplitudes} are valid for any axisymmetric initial conditions described by the functions $f(r,z)$ and $g(r,z)$ in Equation~(\ref{init_cond}). Here we consider $g(r,z)=0$ (i.e., the perturbed system is initially at rest) and $f(r,z)$ given by

\begin{equation}\label{f_r}
f(r,z) = \xi_0\psi(r)\exp(-z^2/\Delta^2),
\end{equation}

\noindent
with

\begin{equation}\label{init_psi}
\psi(r) = \left\{\begin{array}{cl} r/a, & r \leq a, \vspace*{2mm}\\ a/r , & r \geq a.
\end{array}\right. 
\end{equation}

\noindent
The parameter $\Delta$ determines the length of the initial tube disturbance. In addition, $\xi_0$ is the maximum radial displacement of the tube boundary at $t=0$. In our calculations we consider various values of $\Delta$.

The Fourier transforms of $f(r,z)$ and $g(r,z)$ with respect to $z$ are

\begin{align}\label{f_r_k}
\begin{array}{l}
\tilde f(r,k)=\xi_0\sqrt{\pi}\Delta \psi(r) \exp(-\Delta^2k^2/4), \\ [2ex]
\tilde g(r,k)=0,
\end{array}
\end{align}

\noindent and so the following formulae for $\mathcal{N}^\pm_j(k)$ and $\mathcal{N}^\pm_\omega(k)$ can be derived

\begin{align}\label{A_numer-main-2}
\mathcal{N}^\pm_j(k) & = -a\xi_0\sqrt{\pi}\Delta\omega_j 
   \exp(-\Delta^2k^2/4) K_0(\kappae a) \nonumber \\
& \times \bigg\{\frac{\rhoi J_2(\ki a)}{\ki^2} - \frac{\rhoe J_0(\ki a)}{\kappae^2}\bigg\},
\end{align}

\begin{align}\label{A_numer_om-main-2}
\mathcal{N}^\pm_\omega(k) &=
   -a^3\xi_0\sqrt{\pi}\Delta\omega \exp(-\Delta^2k^2/4)
   \bigg\{\frac{\rhoi \ke^2}{\ki^2} J_2(\ki a) \nonumber \\
& + \rhoe \left[C_J J_0(\ke a)+C_Y Y_0(\ke a)\right]\bigg\}.
\end{align}

\noindent Therefore, analytical expressions for all the integrals in the amplitudes $A_j^\pm(k)$ and $A_\omega^\pm(k)$ have been obtained. Note that if $g(r,z)=0$ then $A_j^+(k)=A_j^-(k)$ and $A_\omega^+(k)=A_\omega^-(k)$.

\section{RESULTS}\label{sect_results}

The procedure to obtain the perturbations with the help of Equations~(\ref{expand_eigen-main}) and (\ref{Four_trans_inv-main}) is described in Section~5.4 of Paper~I and thus is not repeated here. In this work we concentrate on the radial displacement at a point on the magnetic tube boundary ($r=a$). The dimensionless results only depend on two parameters, namely the density ratio $\rhoi/\rhoe$ (or equivalently the Alfv\'en speed ratio $\vai/\vae$) and $\Delta$. This parameter determines the length of the initial perturbation, which is of the order of $4\Delta$; see Equation~(\ref{f_r}). We use the parameter values $\rhoi/\rhoe=4,9$ and $\Delta=a,2a$. If dimensional results are needed, then two more parameters are required: the tube radius, $a$, and Alfv\'en speed, $\vai$.

\subsection{Contribution of Proper and Improper Modes}\label{subsubsect_t=0}

\begin{figure}[ht!]
    \scriptsize{(a)}\hspace{-10ex}
    \includegraphics[width=0.35\textwidth,angle=-90]{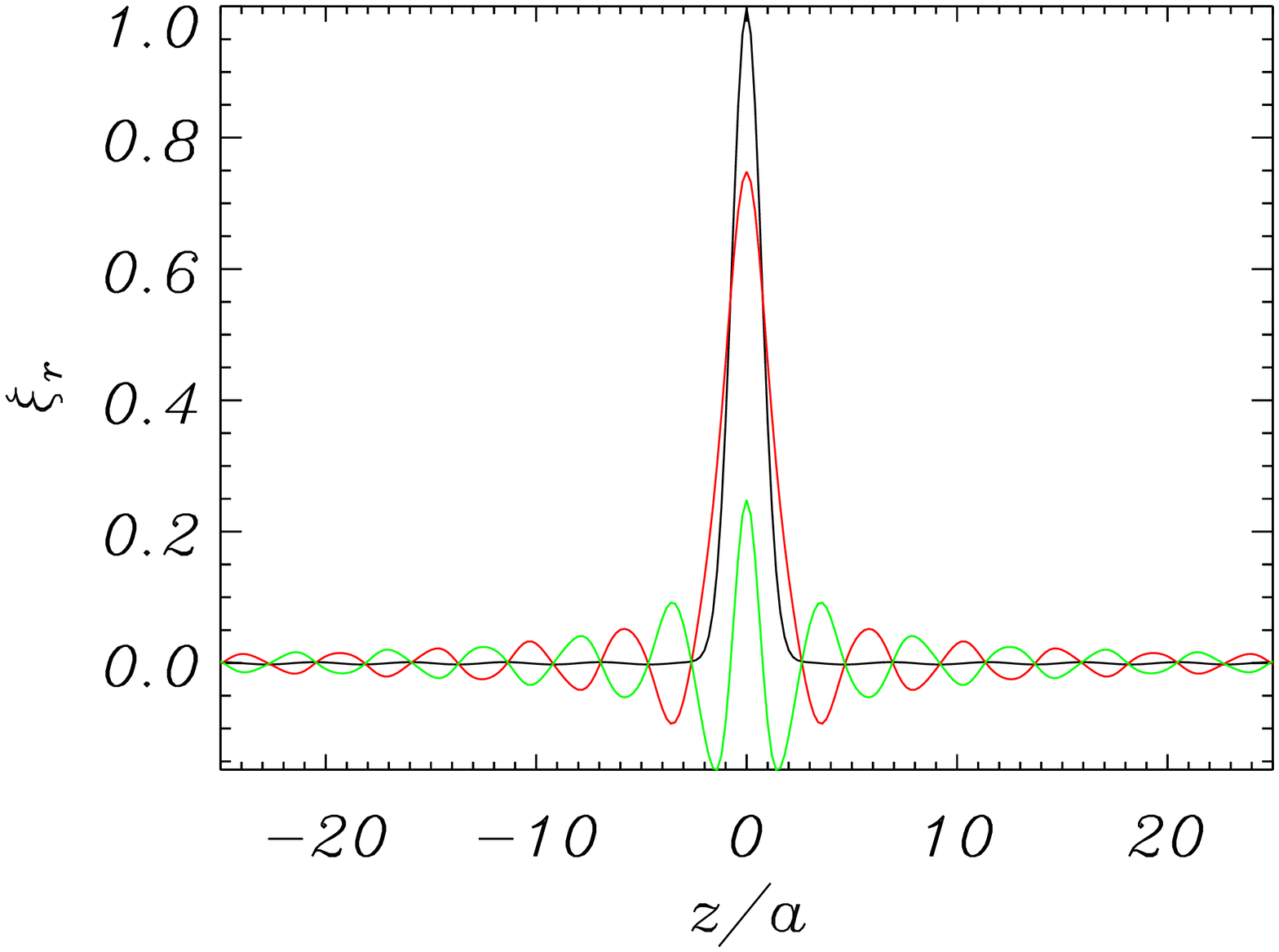} \\
    \scriptsize{(b)}\hspace{-10ex}
    \includegraphics[width=0.35\textwidth,angle=-90]{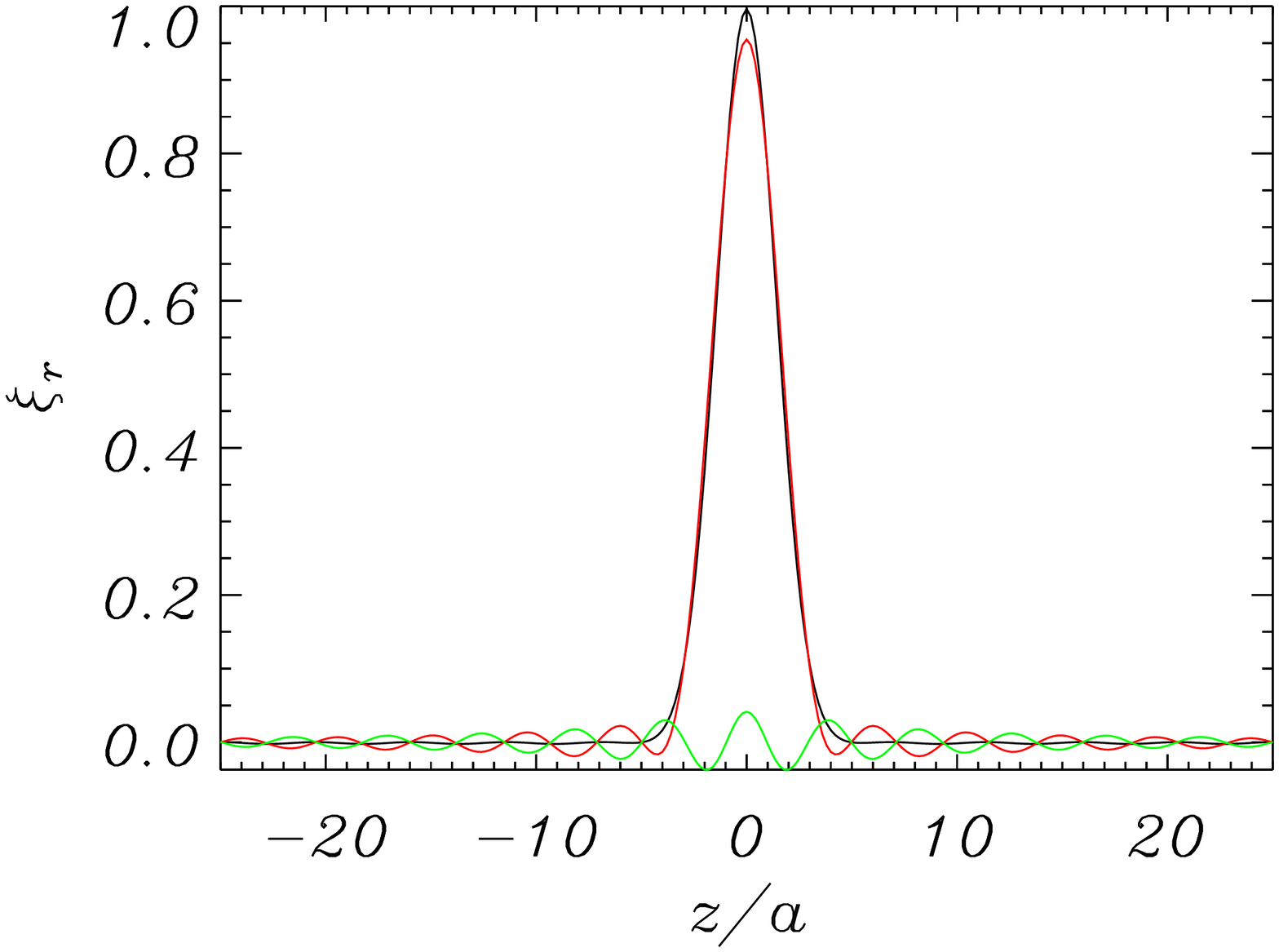} \\
    \scriptsize{(c)}\hspace{-10ex}
    \includegraphics[width=0.35\textwidth,angle=-90]{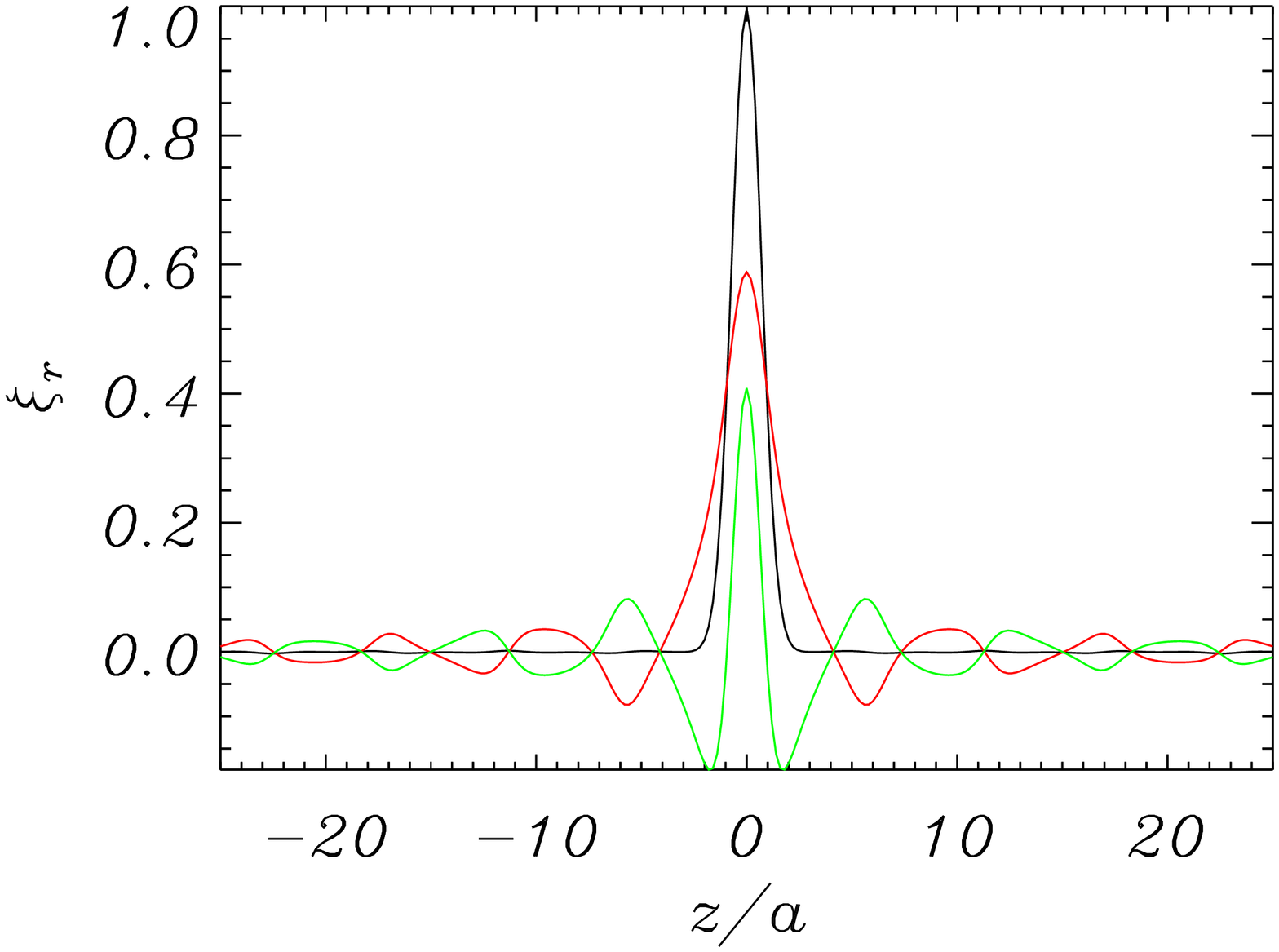}
  \caption{Initial shape of the magnetic tube boundary, given by Equations~(\ref{f_r}) and (\ref{init_psi}) with $r=a$. The radial displacement of the tube boundary is plotted at $t=0$ as a function of the distance along the magnetic tube (black line). The contributions of proper and improper modes are shown as green and red lines, respectively. (a) Density ratio $\rhoi/\rhoe=4$ and $\Delta=a$; (b) $\rhoi/\rhoe=4$ and $\Delta=2a$; (b) $\rhoi/\rhoe=9$ and $\Delta=a$. According to Equation~(\ref{f_r}) the length of the initial disturbance is of the order of $4\Delta$.}
  \label{fig_t=0}
\end{figure}

We first consider the radial displacement at the magnetic tube boundary for $t=0$. This is obtained from Equations~(\ref{Four_trans_inv-main}) and (\ref{expand_eigen-main}), where the contribution of proper and improper modes are computed separately. The black line of Figure~\ref{fig_t=0} shows that this numerically computed $\xi_r$ reproduces the Gaussian initial condition of Equations~(\ref{init_cond}) and (\ref{f_r}). In addition, Figure~\ref{fig_t=0} also reveals that improper modes (red line) have a large contribution to the initial wave form. For the considered parameter values, this contribution is larger than that of proper modes (green line). The reason for the large amplitude of improper modes can be found in the wavenumber cut-off of the fundamental fast sausage eigenmode, that does not exist for $|k|\leq k_{c1}$. Its overtones have even larger wavenumber cut-offs. Then, given the initial disturbance of Equation~(\ref{f_r}) and its Fourier transform of Equation~(\ref{f_r_k}), that peaks at $k=0$, the contributions to $\tilde f(r,k)$ for $|k|\leq k_{c1}$ cannot be accounted for by proper modes. This implies that improper modes must have an important contribution to the initial amplitude of the perturbation and, therefore, to the posterior evolution of the wave train propagation. Another significant feature of Figure~\ref{fig_t=0} is that the contribution of both proper and improper modes extends well beyond the length of the tube affected by the initial disturbance. The two contributions display spatial variations that added together yield the Gaussian shape of Equation~(\ref{f_r_k}).

It is worth noticing that improper eigenmodes are more relevant for the present axisymmetric disturbance than for the transverse disturbance: compare Figure~\ref{fig_t=0} and Figures~4(a) and (b) of Paper~I. The main difference between fast kink and sausage proper modes is that the fundamental proper kink mode exists for all wavenumbers, so that it can account for the main part of the function $\tilde f(r,k)$. This fact lowers the importance of improper modes in the case of a transverse disturbance and also has a second effect: contrary to the case studied here, at $t=0$ proper and improper modes are confined to the vicinity of the initial perturbation, such as revealed by Figures~4(a) and (b) of Paper~I.

\begin{figure}[ht!]
    \scriptsize{(a)}\hspace{-10ex}
    \includegraphics[width=0.35\textwidth,angle=-90]{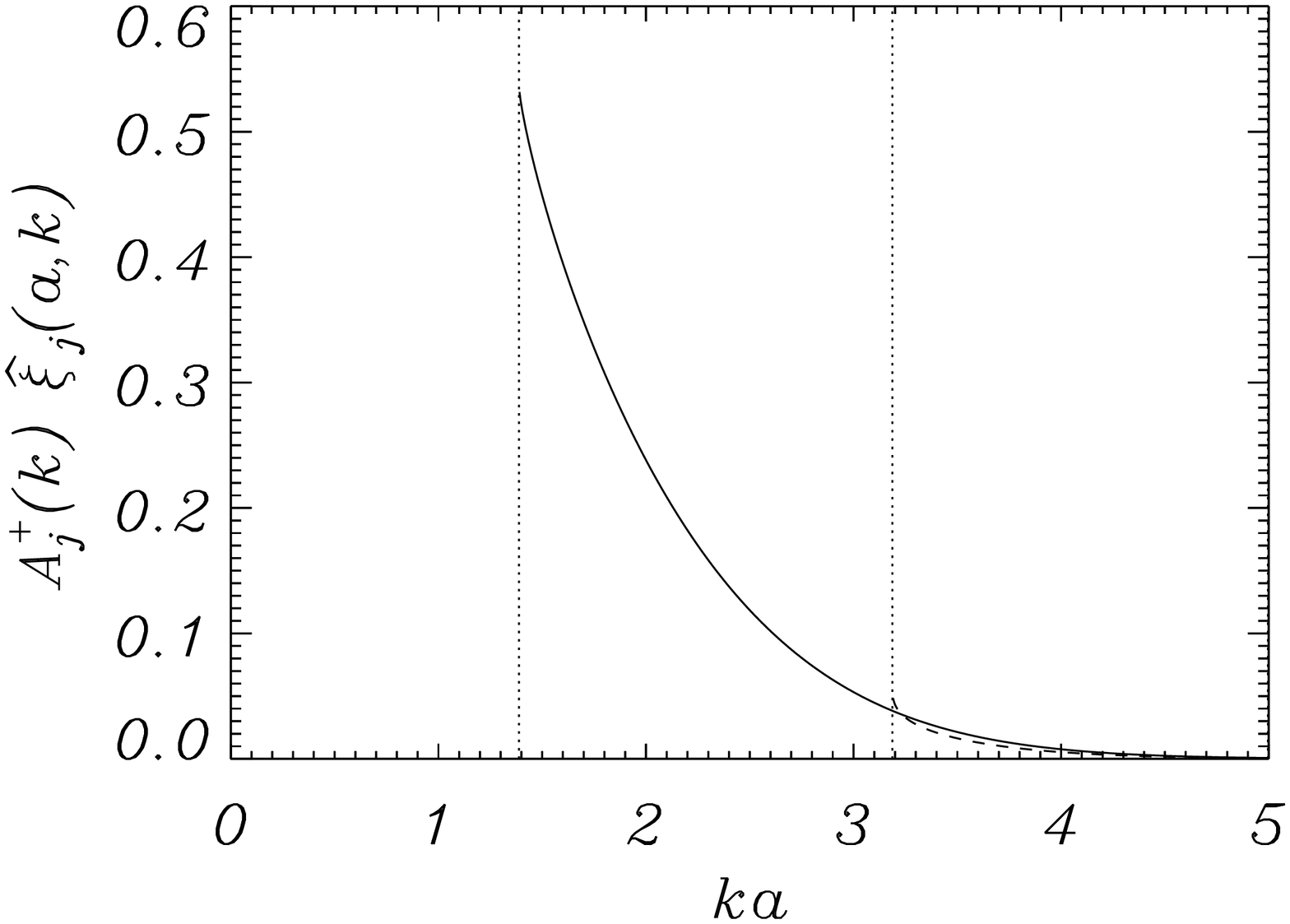} \\
    \scriptsize{(b)}\hspace{-10ex}
    \includegraphics[width=0.35\textwidth,angle=-90]{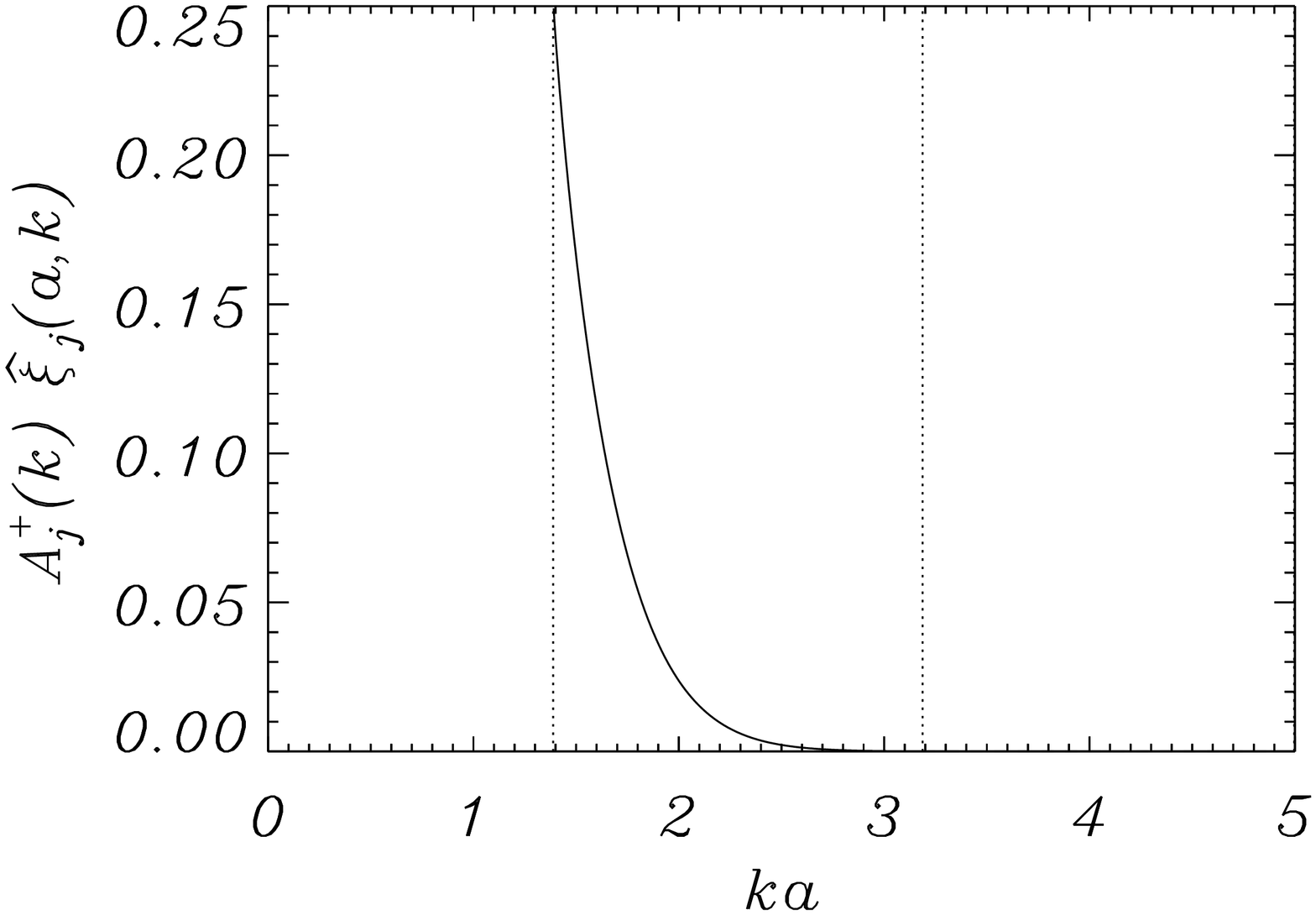} \\
    \scriptsize{(c)}\hspace{-10ex}
    \includegraphics[width=0.35\textwidth,angle=-90]{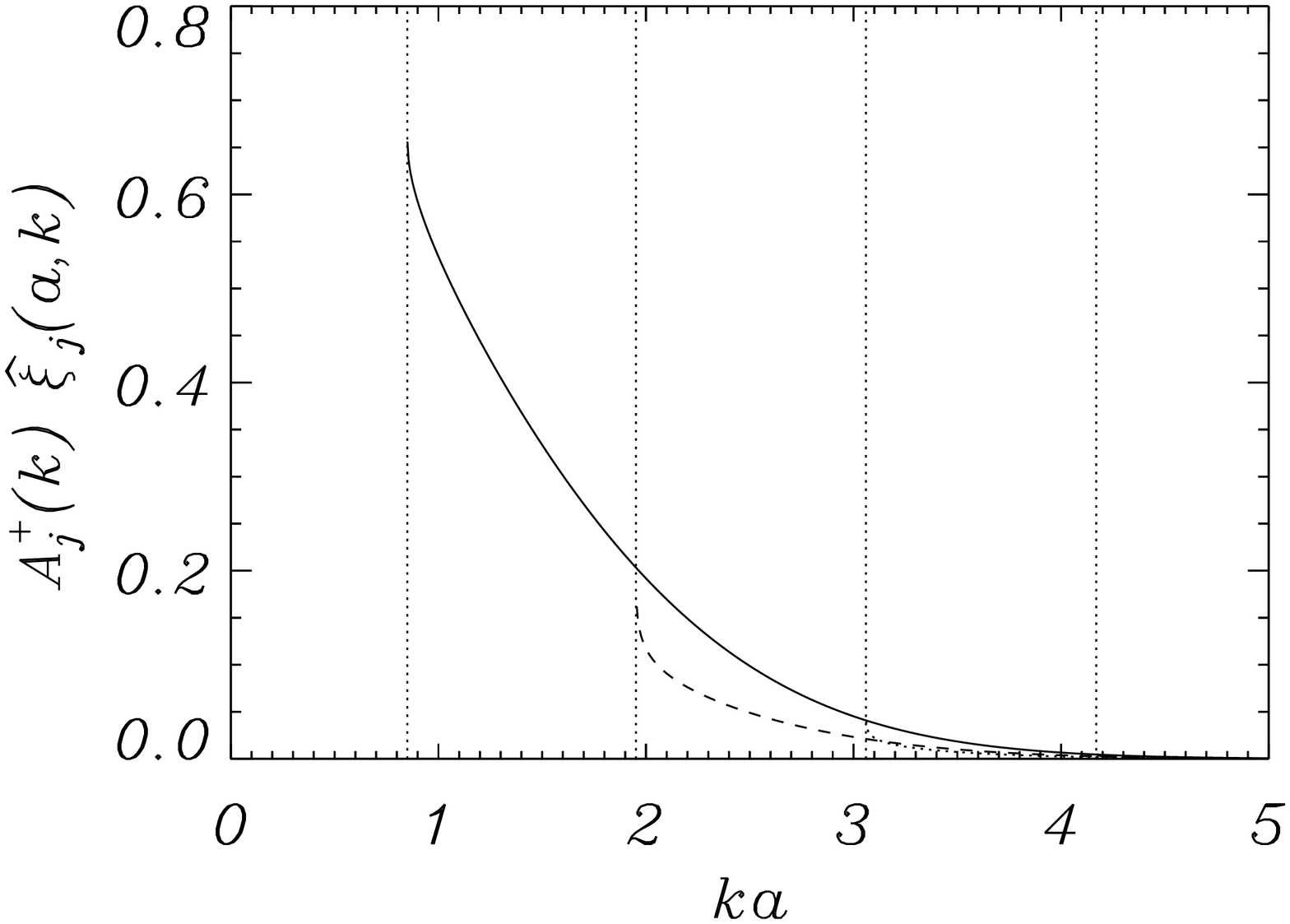}
  \caption{Proper mode contribution to the radial displacement as a function of the longitudinal wavenumber. When inserted in Equation~(\ref{expand_eigen-main}), the product $A_j^+(k)\hat\xi_j(a,k)$ determines the contribution of each proper mode to the radial displacement for a fixed~$k$. (a) $\rhoi/\rhoe=4$ and $\Delta=a$; (b) $\rhoi/\rhoe=4$ and $\Delta=2a$; (c) $\rhoi/\rhoe=9$ and $\Delta=a$. Solid, dashed, and dotted lines correspond to the fundamental mode and its first two overtones. The vertical dotted lines mark the wavenumber cut-offs.}
  \label{fig_amplr}
\end{figure}

From Figure~\ref{fig_t=0} we also see that the contribution of improper modes increases when either $\Delta$ is increased or $\rhoi/\rhoe$ is decreased. We next explain this behavior. An increase of $\Delta$ corresponds to a longer initial perturbation, whose Fourier transform becomes more concentrated around $k=0$. Since the fundamental fast sausage mode only exists for $|k|>k_{c1}$, an increase in $\Delta$ results in a decrease of $\tilde f(r,k)$ for $|k|>k_{c1}$, so that the contribution of proper modes becomes smaller. This is corroborated by Figures~\ref{fig_amplr}(a) and (b), where the contribution of proper modes to the amplitude of $\tilde\xi_r$ at $r=a$ is shown for two values of $\Delta$. In addition, the role of the density ratio is to modify the cut-off wavenumbers, in such a way that a decrease in $\rhoi/\rhoe$ leads to a larger wavenumber cut-off (see Figure~\ref{fig_dr_over}). This then yields a smaller contribution of proper modes: compare Figures~\ref{fig_amplr}(a) and (c).

Figure~\ref{fig_amplr} also provides information about the relative importance of proper mode overtones. Such as expected, the largest amplitude of all proper modes corresponds to the fundamental fast sausage mode, whereas its radial overtones possess smaller amplitudes. By comparing panels (a) and (b) we see that longer initial perturbations result in a negligible amplitude of all radial overtones. We must mention that in Figures~\ref{fig_amplr}(a) and (b) only the fundamental mode and its first overtone are present (solid and dashed lines), although in panel (b) the first overtone has negligible amplitude, so its curve is invisible. Furthermore, in Figure~\ref{fig_amplr}(c) the dashed and dotted curves overlap for $3\lesssim ka\lesssim 4$ and may seem to be just one.

\subsection{Propagation of Wave Trains}

\begin{figure*}[ht!]
  \centerline{
    \scriptsize{(a)}\hspace{-10ex}
    \includegraphics[width=0.35\textwidth,angle=-90]{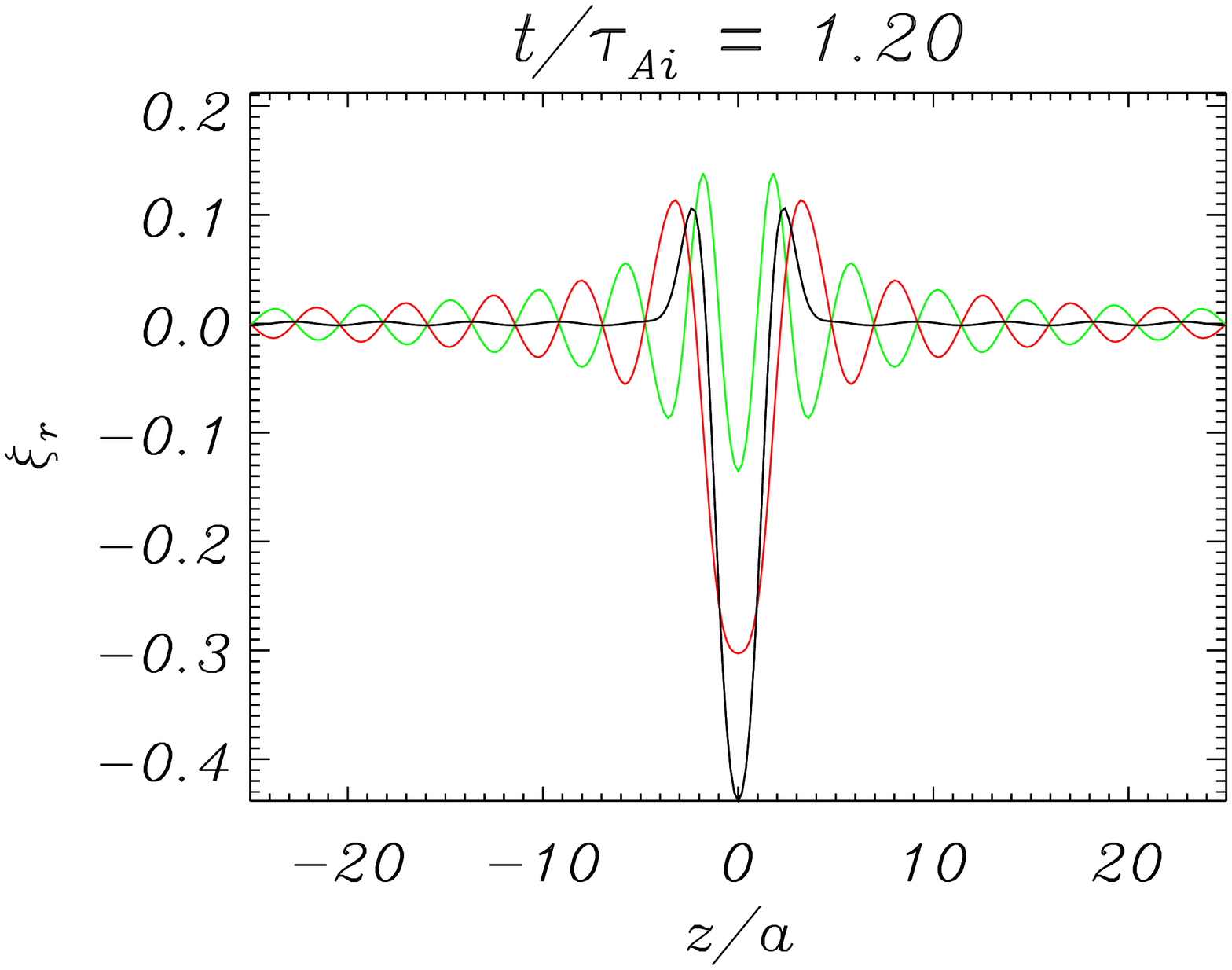} \\
    \scriptsize{(b)}\hspace{-10ex}
    \includegraphics[width=0.35\textwidth,angle=-90]{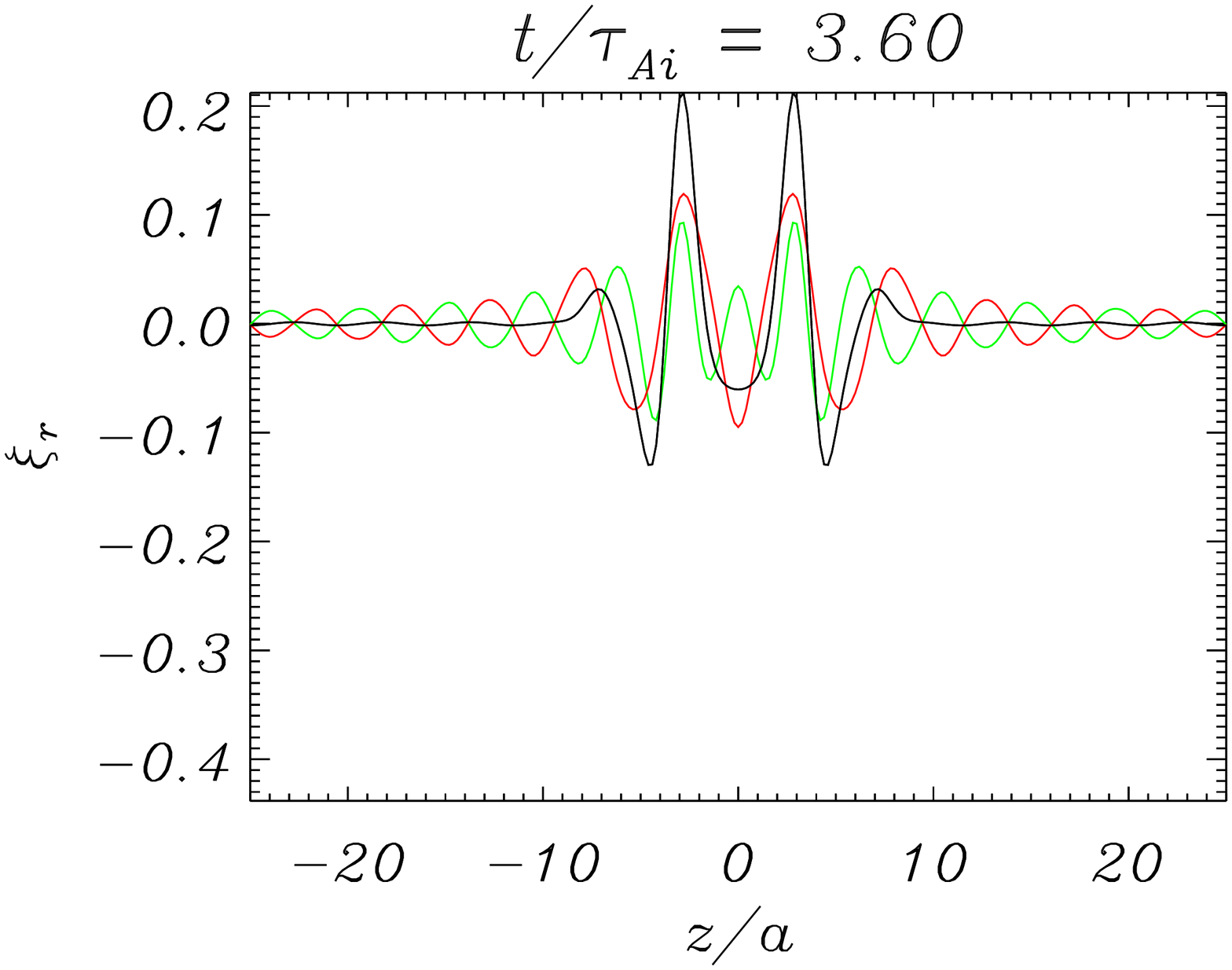} \\
  }
  \centerline{
    \scriptsize{(c)}\hspace{-10ex}
    \includegraphics[width=0.35\textwidth,angle=-90]{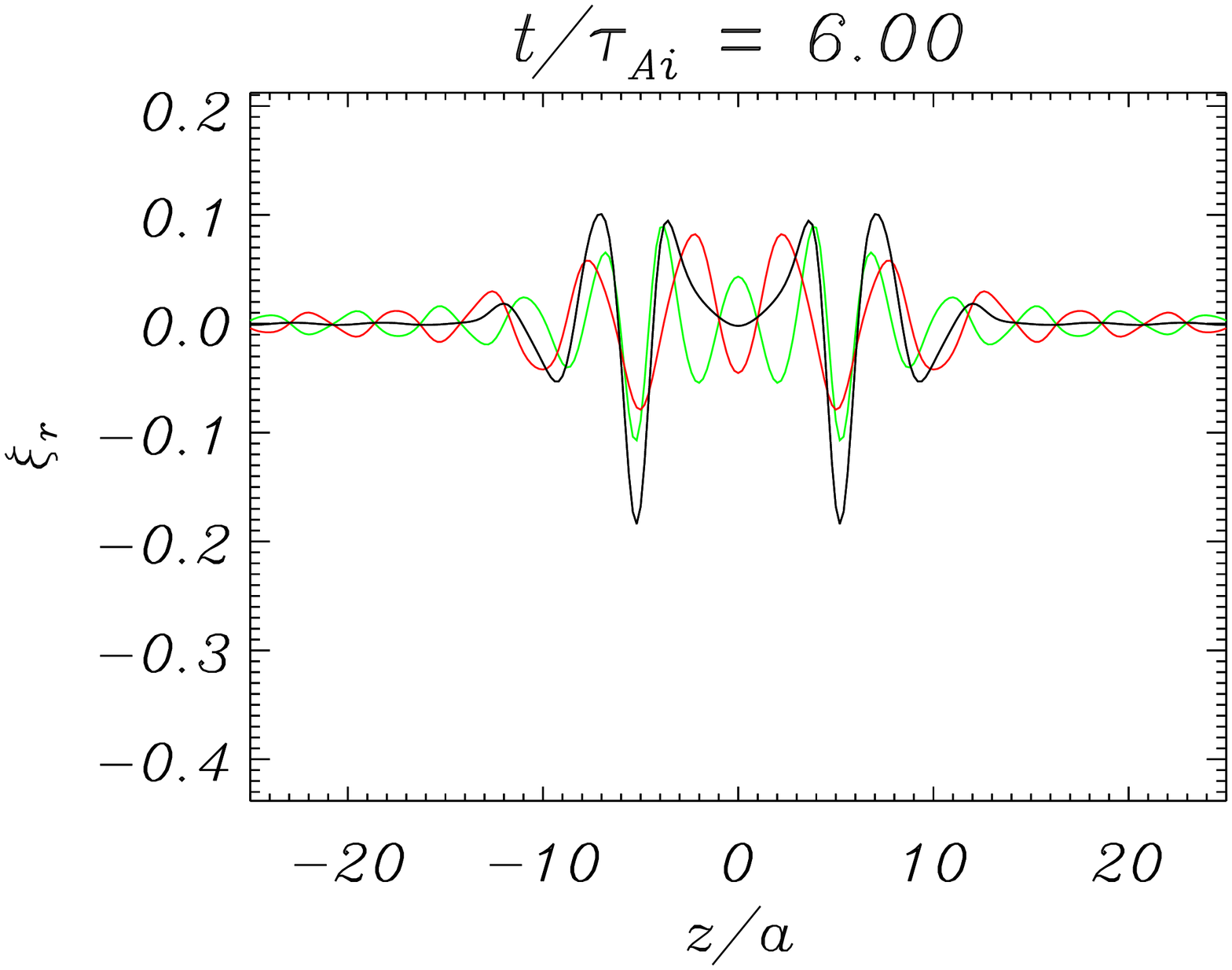} \\
    \scriptsize{(d)}\hspace{-10ex}
    \includegraphics[width=0.35\textwidth,angle=-90]{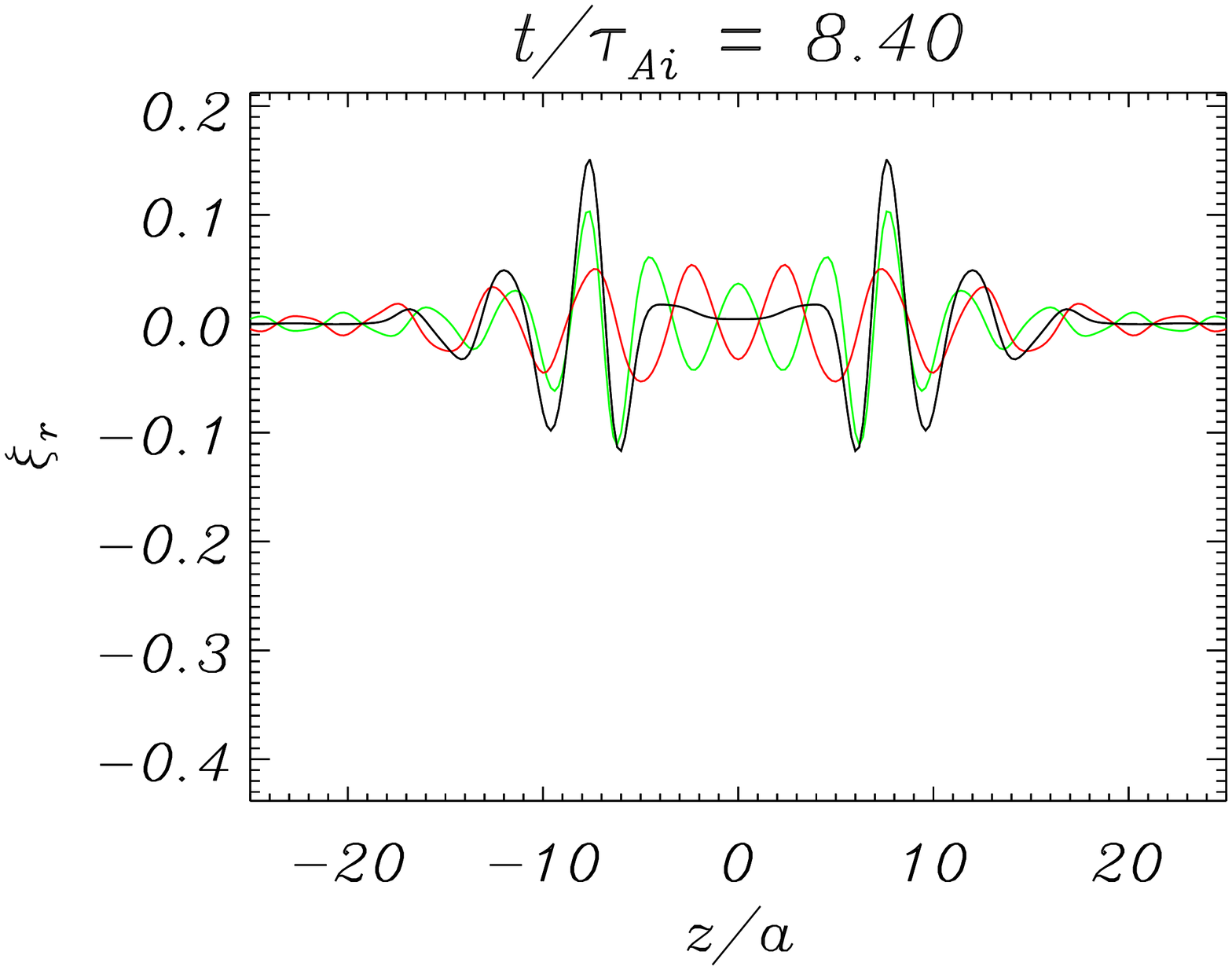}
  }
  \caption{Shape of the magnetic tube boundary for several times (shown at the top of each frame). The radial displacement at $r=a$ is plotted as a function of $z$. This figure shows the early evolution of the initial pulse of Figure~\ref{fig_t=0}(a). Black lines correspond to $\xi_r$, while green and red lines correspond to the contributions of proper and improper modes, respectively. The density ratio is $\rhoi/\rhoe=4$ and the length of the initial disturbance is of the order of $4a$. \newline (An animation of this figure is available.)}
  \label{fig_vae2_delta1_short_t}
\end{figure*}

Next, we concentrate in the evolution of the pulse of Figure~\ref{fig_t=0}(a). This initial perturbation splits in two identical wave trains that propagate symmetrically in both directions along the magnetic tube. The early stages of the propagating wave trains are presented in Figure~\ref{fig_vae2_delta1_short_t} and Animation~1. Figures~\ref{fig_vae2_delta1_short_t}(a)--(c) correspond to the separation of the two wave trains, whereas in Figure~\ref{fig_vae2_delta1_short_t}(d) these two features are fully formed. The strong dispersion of the axisymmetric perturbation can be well appreciated in this last panel: both wave packets have developed some 6 maxima and minima and their length is more than 4 times that of the perturbation at $t=0$. Proper and improper modes contribute with similar amplitudes to the total signal represented in Figure~\ref{fig_vae2_delta1_short_t} and Animation~1. A secondary effect of wave dispersion is the strong reduction of the wave packets amplitude. In the absence of dispersion, the two wave packets of Figure~\ref{fig_vae2_delta1_short_t}(d) would have amplitude 0.5, whereas their actual peak-to-peak amplitude is under 0.3.

When discussing Figure~\ref{fig_t=0} we explained that, when considered separately, the contributions of proper and improper modes affect a section of the magnetic tube much longer than that of the initial perturbation. At $t=0$ these two contributions simply cancel out in the parts of the magnetic tube where the initial excitation is negligible. This situation also takes place for $t>0$: although proper and improper mode contributions disturb a large part of the magnetic tube, their sum is concentrated in a smaller section. Thus, ahead of the propagating wave fronts, where the initial perturbation has not yet arrived, these contributions are spatially out of phase and cancel out. This is also found at $-2\lesssim z/a\lesssim 2$ of Figure~\ref{fig_vae2_delta1_short_t}(d) because the two wave packets have moved beyond this section of the magnetic tube, which is barely disturbed.

\begin{figure}[ht!]
    \scriptsize{(a)}\hspace{-10ex}
    \includegraphics[width=0.35\textwidth,angle=-90]{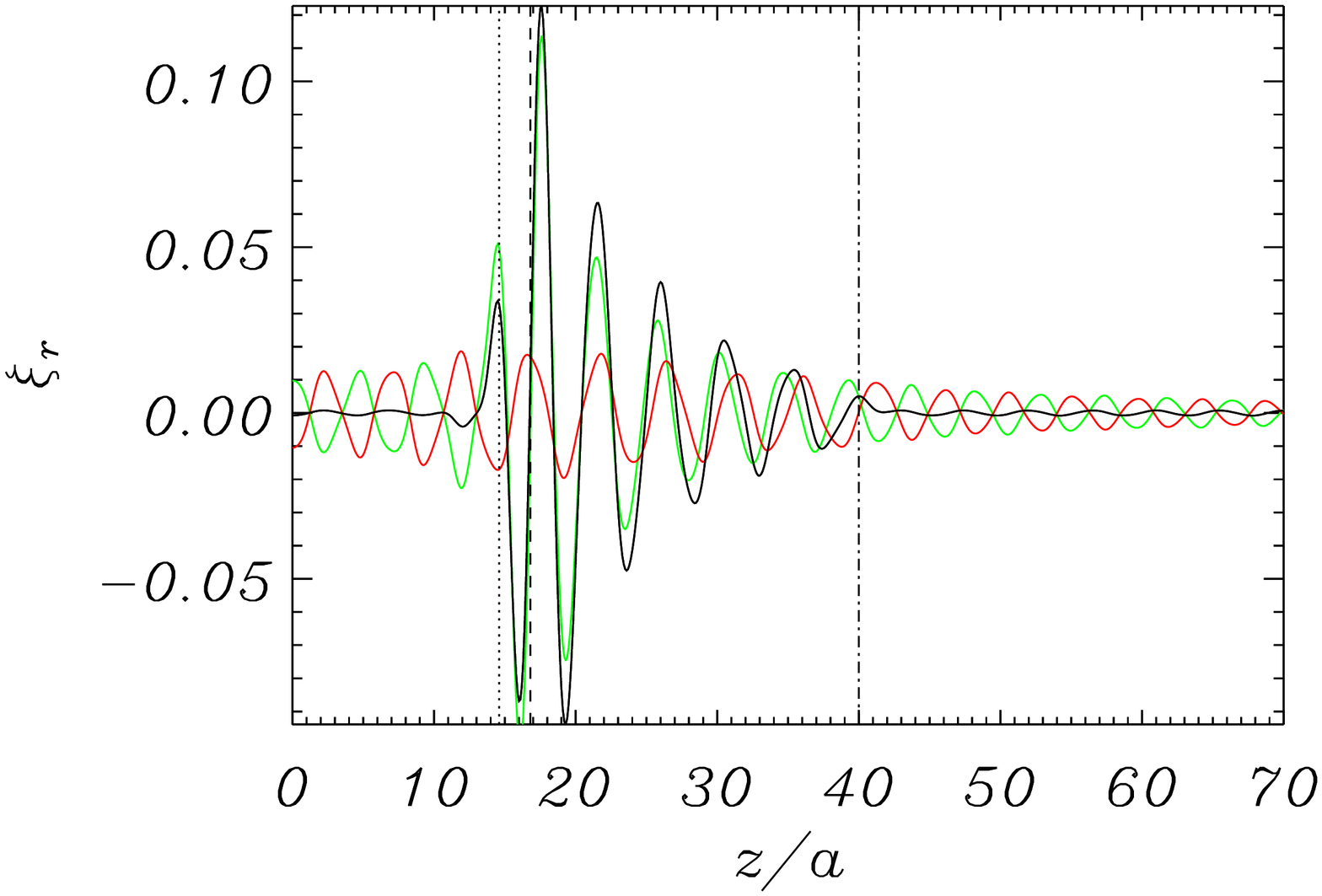} \\
    \scriptsize{(b)}\hspace{-10ex}
    \includegraphics[width=0.35\textwidth,angle=-90]{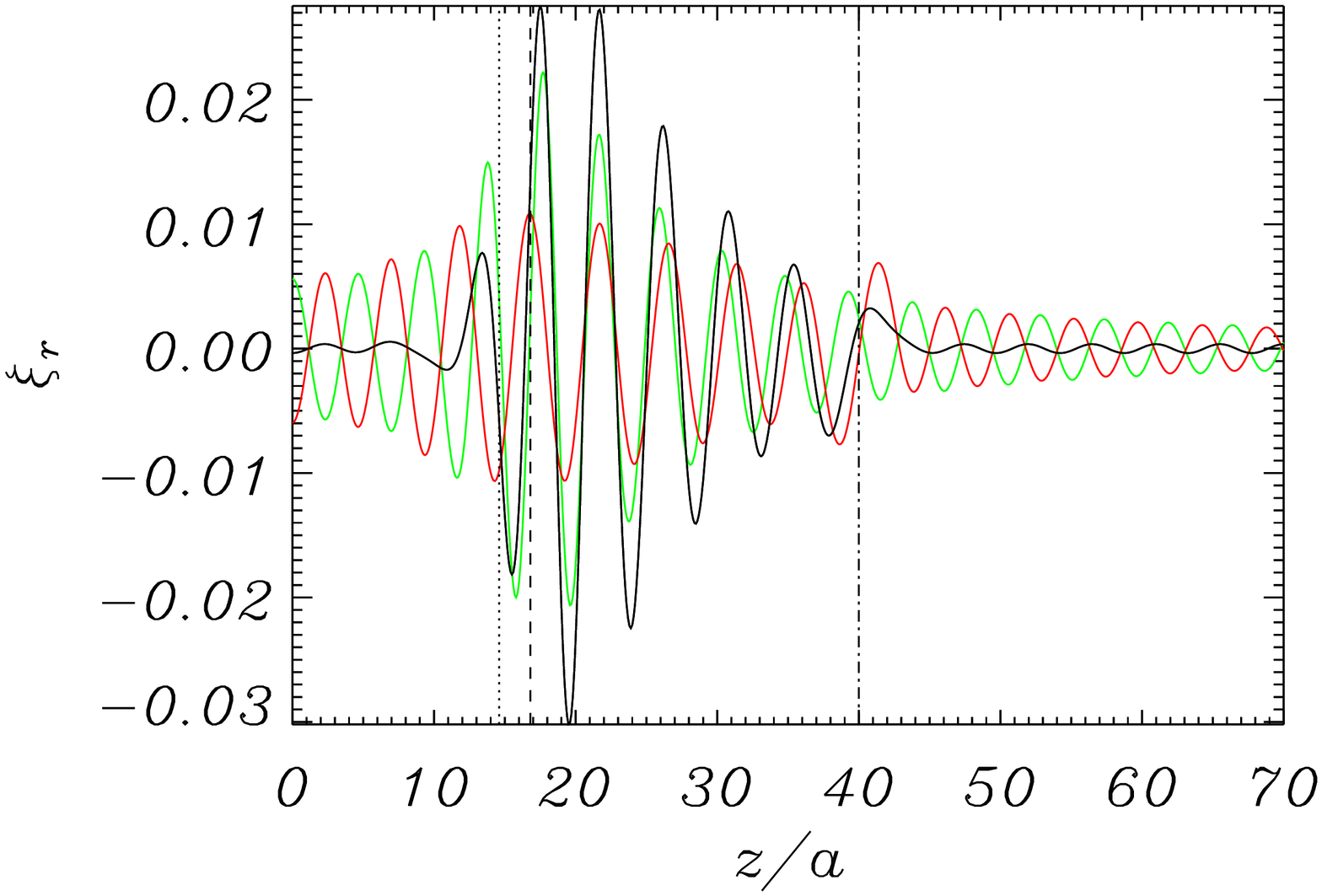} \\
    \scriptsize{(c)}\hspace{-10ex}
    \includegraphics[width=0.35\textwidth,angle=-90]{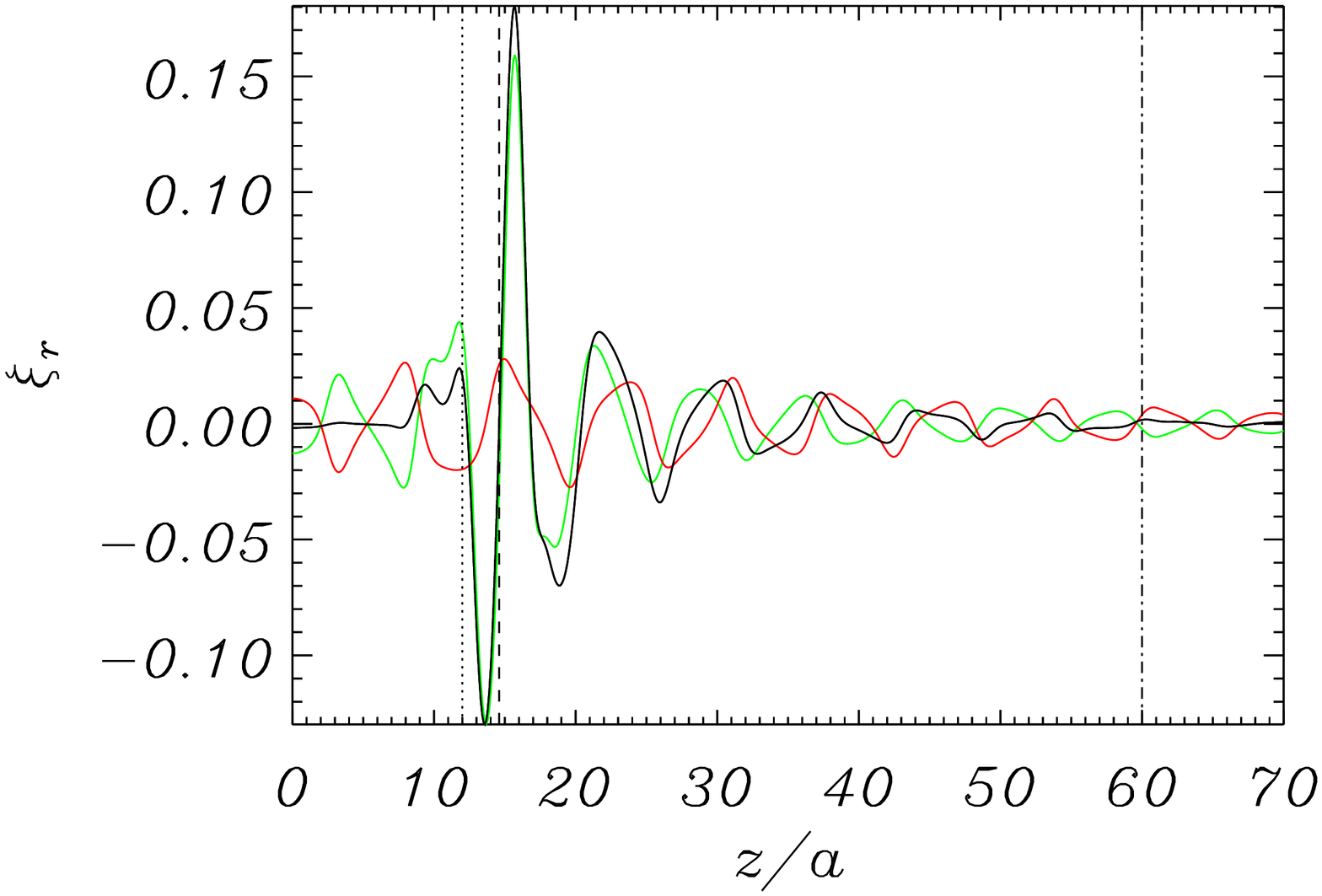}
  \caption{Shape of the magnetic tube boundary at $t=20\tauAi$. The three panels correspond to the wave trains caused by the initial disturbances of Figure~\ref{fig_t=0}, with the same meaning of line colors. The vertical lines give the position of a point initially at $z=0$ and traveling at a constant speed equal to the maximum group velocity of proper modes (dash-dotted line), the minimum group velocity of the fundamental proper mode (dashed line), and the minimum group velocity of the first proper mode overtone (dotted line).}
  \label{fig_longer_t}
\end{figure}

The green and red curves of Figure~\ref{fig_vae2_delta1_short_t} reveal the separate behavior of proper and improper modes. The first ones propagate with group velocity ranging from $\sim 0.84\vai$ (see the minimum of the black line in Figure~\ref{fig_dr_over}(c)) to $\vae=2\vai$. Because of this variation of group velocity the green curve displays some dispersion, that becomes more clear for later times (this issue will be discussed below with the help of Figure~\ref{fig_longer_t}). On the other hand, improper modes propagate at their own phase speed and so are subject to phase mixing, that leads to stronger dispersion. This dispersive behavior in turn causes a severe amplitude decrease that is clearly visible when comparing the red lines of Figures~\ref{fig_vae2_delta1_short_t}(a) and (d). The asymptotic behavior of improper modes for large $t$ was studied in Section~5.3 of Paper~I, that deals with a transverse initial perturbation. It was found that the contribution of improper modes decays as $1/t$. The same applies to the present axisymmetric perturbation.

At $t=20\tauAi$ the perturbation of Figure~\ref{fig_t=0}(a), whose early development is presented in Figure~\ref{fig_vae2_delta1_short_t}, is spread over a long section of the tube of length $\sim 30a$ (compare this value with the initial length of the perturbation $\sim 4a$); see Figure~\ref{fig_longer_t}(a). Here the total perturbation of the radial displacement and the contributions of proper and improper modes are represented with the usual colors (black, green, and red, respectively). The vertical lines denote the position of a point initially at $z=0$ and moving at specific velocities: the dash-dotted, dashed, and dotted lines correspond to the maximum group velocity of proper modes ($2\vai$), the minimum group velocity of the fundamental fast sausage mode ($0.84\vai$), and the minimum group velocity of its first overtone ($0.73\vai$). This range of group velocities determines the length of the wave train at a given time. The wave train shape is controlled by two more features: the dependence $\cg(k)$ of proper modes (black lines in Figure~\ref{fig_dr_over}(c)) and the amplitude of the perturbation (Figure~\ref{fig_amplr}(a)). Wavenumbers slightly above the cut-off occupy the leading part of the wave packet and have the largest amplitudes, but their $\cg$ varies strongly with $k$. Thus, these wavenumbers are strongly dispersed and contribute moderately to the wave train amplitude. On the other hand, wavenumbers $ka\gtrsim 1.6$ have group velocities in a small range around the minimum of $\cg$ and for this reason the maximum amplitude of the wave train is found near the dashed line of Figure~\ref{fig_longer_t}(a). A side effect of the wave dispersion is the decrease of the wave packet amplitude (compare Figures~\ref{fig_vae2_delta1_short_t}(d) and \ref{fig_longer_t}(a)). Again, note the cancellation of proper and improper mode contributions ahead and behind the propagating wave train.

We now investigate the effect of the initial disturbance length and the density ratio on the wave train dispersion. If  the internal to external density ratio is kept fixed and the length of the initial disturbance is doubled (Figure~\ref{fig_longer_t}(b)) the dispersion of the wave train in its constituent frequencies does not change because the group velocity of proper modes is the same. Then, the positions of maxima and minima in Figures~\ref{fig_longer_t}(a) and (b) are the same. Nevertheless, according to Figures~\ref{fig_t=0}(a) and (b) and Figures~\ref{fig_amplr}(a) and (b), a longer initial disturbance corresponds to smaller amplitudes of proper modes and for this reason the amplitude of the wave train is much smaller for $\Delta=2a$ than for $\Delta=a$.

Finally, we consider the case of a magnetic tube with a larger internal to external density contrast ($\rhoi/\rhoe=9$; Figure~\ref{fig_longer_t}(c)). In this case the minimum group velocity of the fundamental fast sausage mode and its first overtone are similar to those for $\rhoi/\rhoe=4$ and this implies that the position of the vertical dashed and dotted lines in Figures~\ref{fig_longer_t}(a) and (c) are similar. But by increasing the density ratio from 4 to 9, the maximum group velocity of proper modes increases from $2\vai$ to $3\vai$ and so the front of the wave train travels considerably faster. Hence, the wave packet extends over a much longer section of the magnetic tube, with length $\sim 50a$. The leading part of the wave train has a small amplitude because of the strong gradient of $\cg(k)$ near the cut-off wavenumber. In addition, the largest amplitude is found at the back of the wave packet, were, again, a wide range of wavenumbers with values of $\cg$ near the minimum group velocity concentrate their contribution.


\subsection{Comparison of Signals Caused by Axisymmetric and Transverse Disturbances}\label{sect_comparison}

\begin{figure*}[ht!]
  \centerline{
    \scriptsize{(a)}\hspace{-10ex}
    \includegraphics[width=0.35\textwidth,angle=-90]{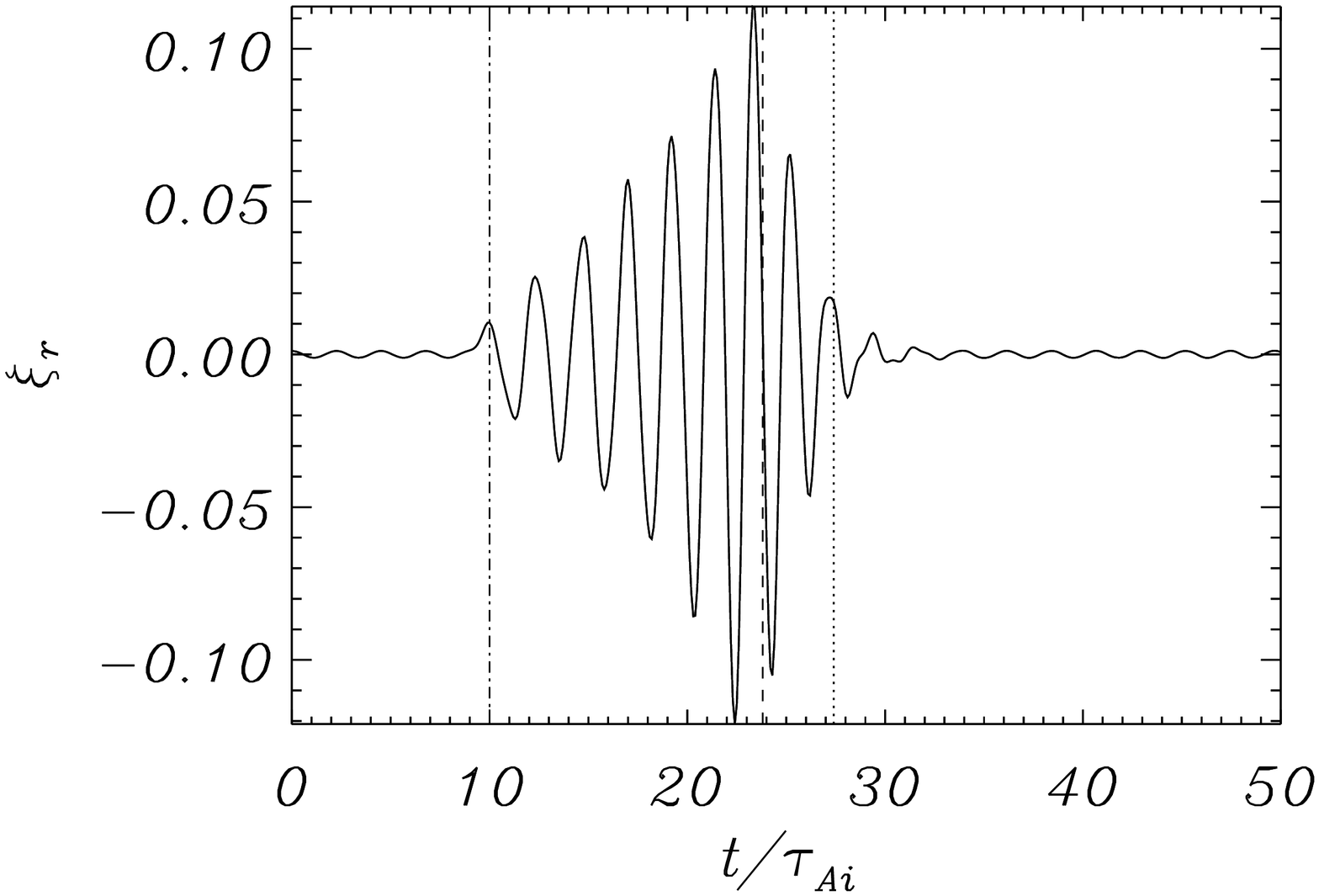} \\
    \scriptsize{(b)}\hspace{-10ex}
    \includegraphics[width=0.35\textwidth,angle=-90]{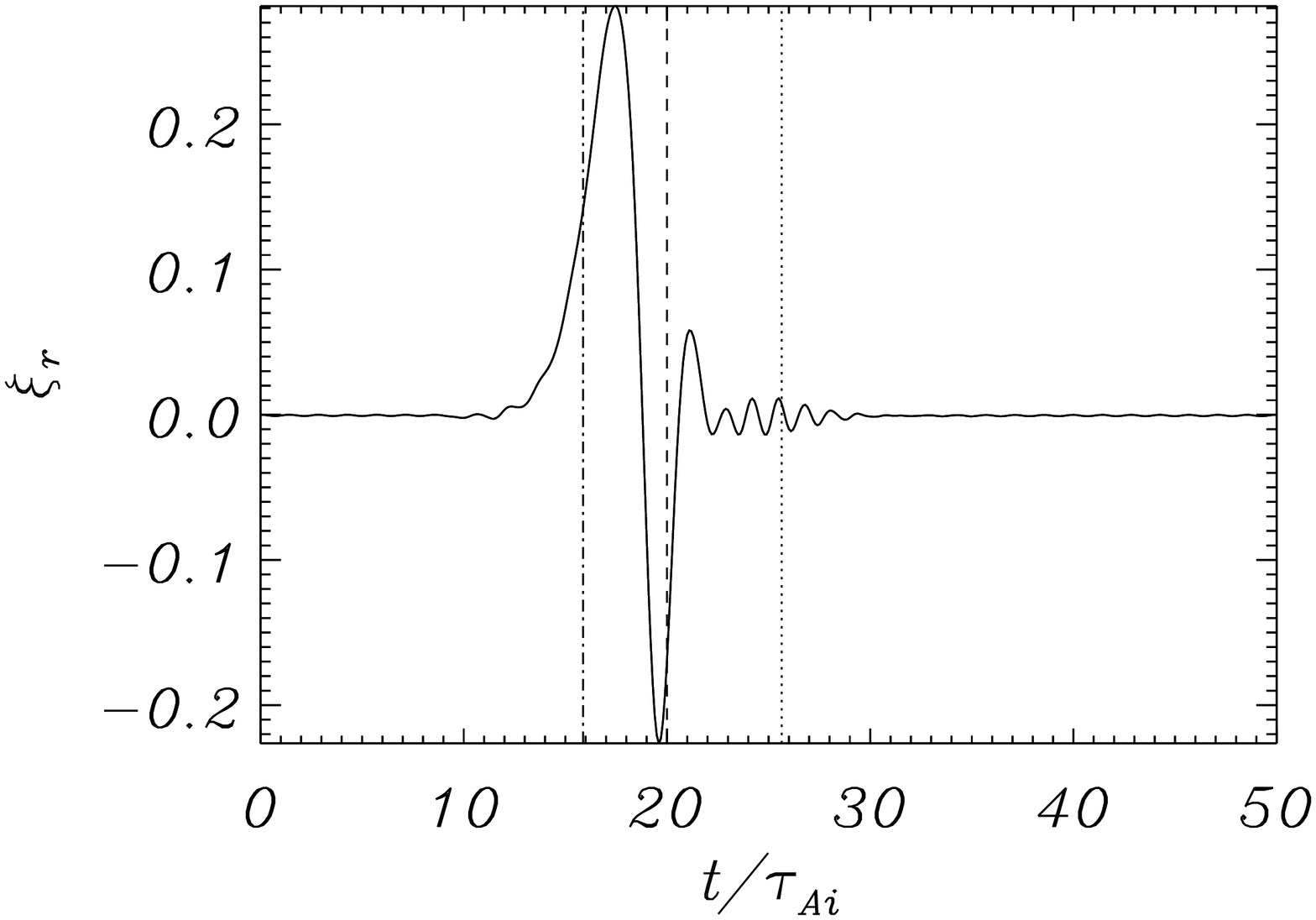} \\
  }
  \centerline{
    \scriptsize{(c)}\hspace{-10ex}
    \includegraphics[width=0.35\textwidth,angle=-90]{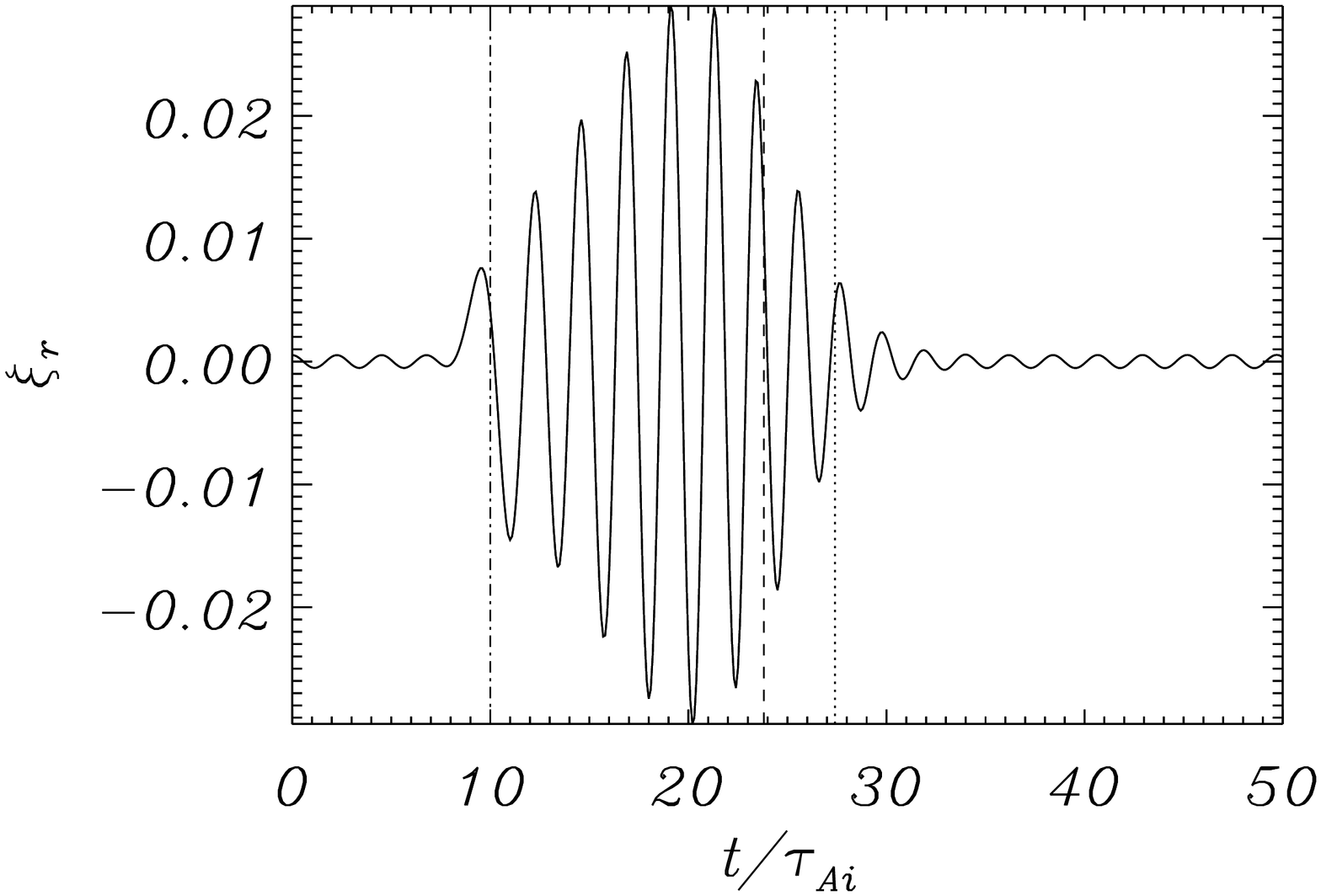} \\
    \scriptsize{(d)}\hspace{-10ex}
    \includegraphics[width=0.35\textwidth,angle=-90]{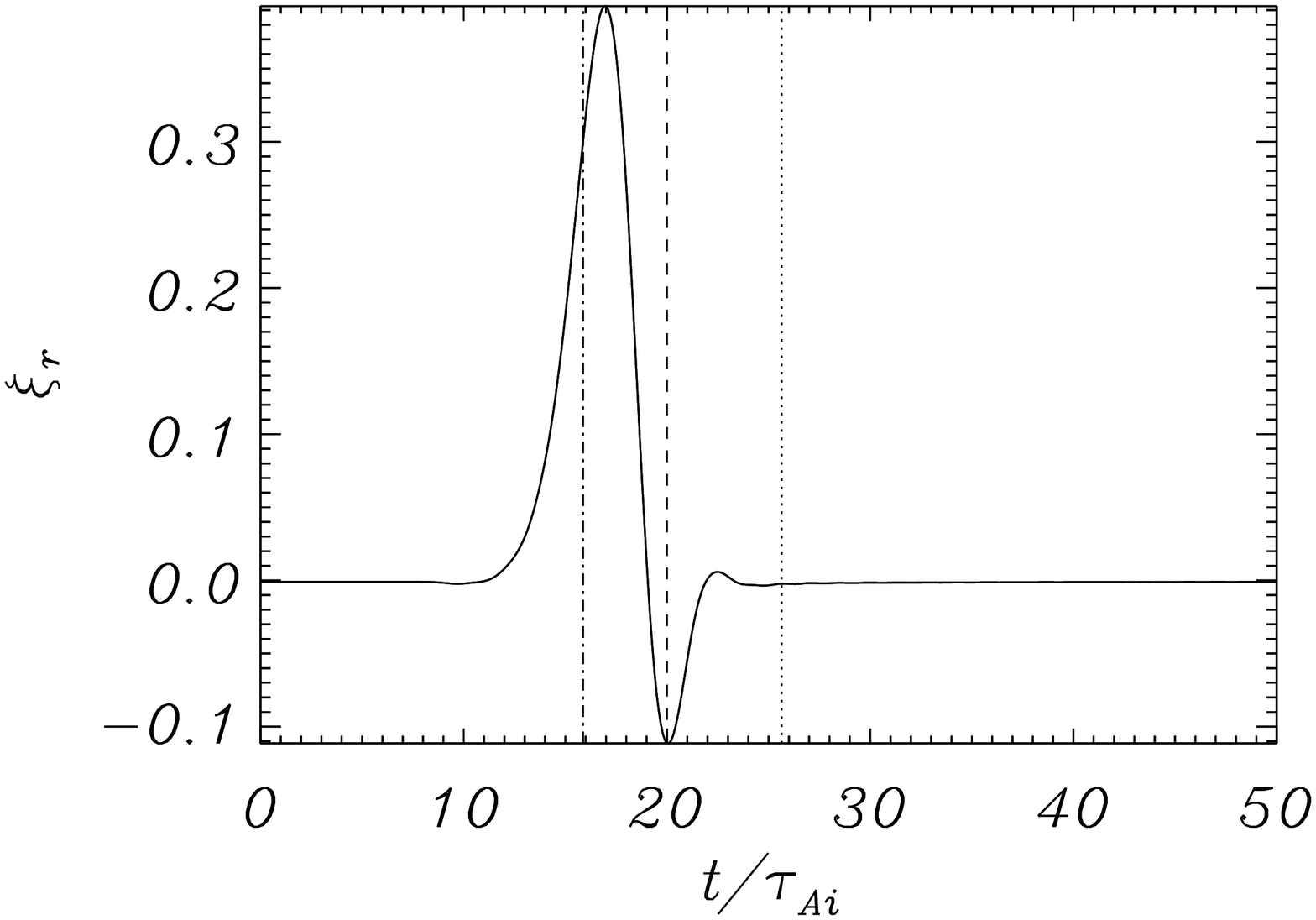} \\
  }
  \centerline{
    \scriptsize{(e)}\hspace{-10ex}
    \includegraphics[width=0.35\textwidth,angle=-90]{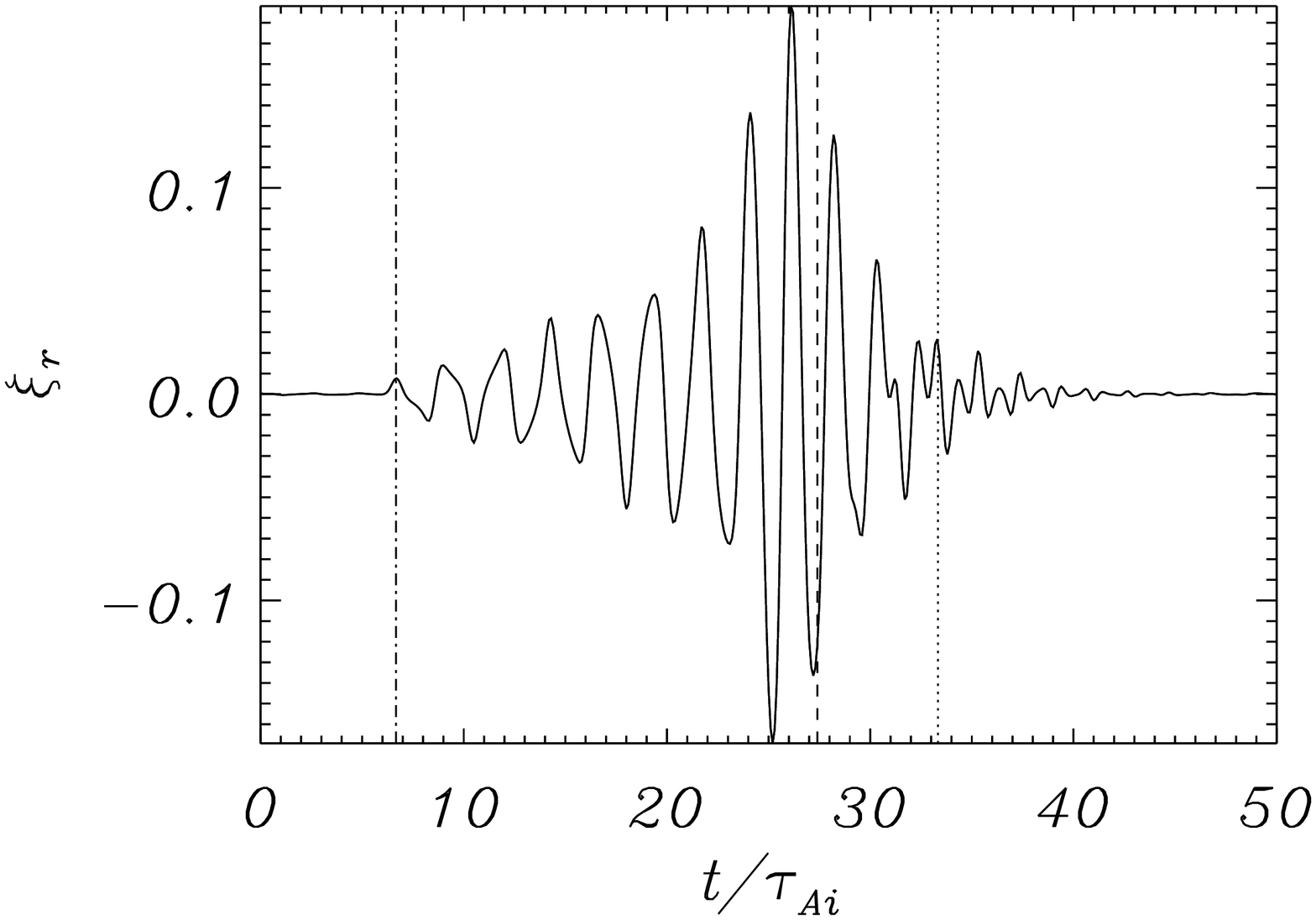} \\
    \scriptsize{(f)}\hspace{-10ex}
    \includegraphics[width=0.35\textwidth,angle=-90]{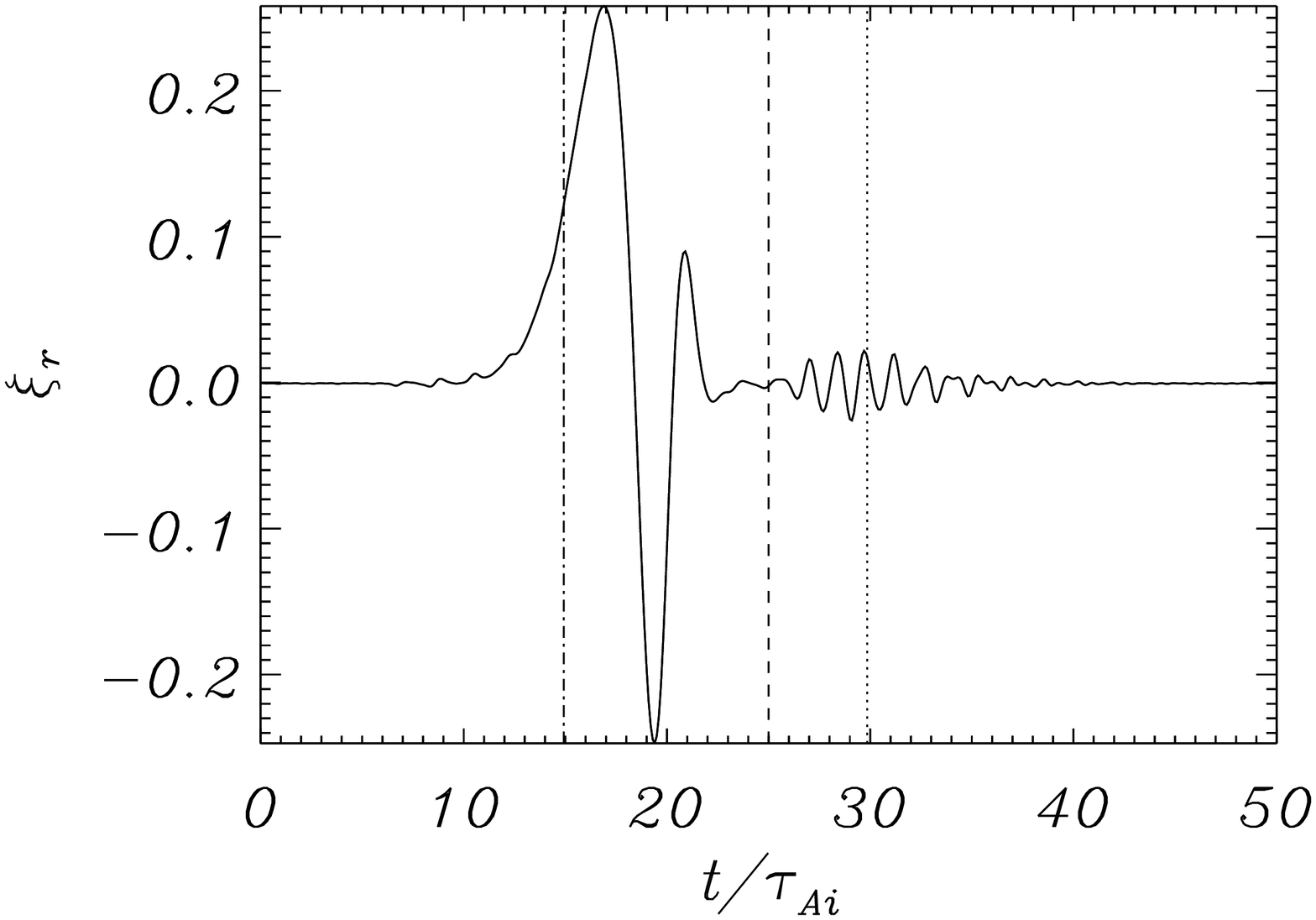}
  }
  \caption{Radial displacement of the magnetic tube boundary at a distance $z=20a$ from the initial perturbation. Left column: axisymmetric (sausage) perturbation; right column: transverse (kink) perturbation. (a) and (b) $\rhoi/\rhoe=4$, $\Delta=a$; (c) and (d) $\rhoi/\rhoe=4$, $\Delta=2a$; (e) and (f) $\rhoi/\rhoe=9$, $\Delta=a$. The vertical lines have the same meaning as in Figure~\ref{fig_longer_t}.}
  \label{fig_comp_kink_sausage}
\end{figure*}

Here we compare the response of the magnetic tube boundary under initial axisymmetric and transverse disturbances, respectively described in this paper and in Paper~I. To make this comparison we consider the wave train passage at a fixed position ($z=20a$) and plot the radial displacement as a function of time (see Figure~\ref{fig_comp_kink_sausage}, in which the left and right columns correspond to axisymmetric and transverse\footnote{This case was studied in Paper~I, although the external variation of $\psi(r)$, which gives the radial profile of the initial perturbation, was different from that used in the present work (Equation~(\ref{init_psi})). In order to make a comparison with the results of Paper~I, the proper and improper mode amplitudes of transverse disturbances are presented in Appendix~\ref{app_kink} for the same $\psi(r)$.} disturbances).

Inspection of Figure~\ref{fig_comp_kink_sausage} reveals the very different dispersive properties of fast sausage and kink waves. The first ones display rapid oscillations with period $\sim 3\tauAi$, whereas the latter have only one full oscillation with larger amplitude and period $\sim 6\tauAi$. This disparate behavior occurs because the fundamental fast kink mode does not have a cut-off wavenumber and its group velocity has a more moderate range of variation than that of its sausage counterpart. The length of the initial perturbation plays a different role in the propagation of axisymmetric and transverse disturbances. In the first case (compare Figures~\ref{fig_comp_kink_sausage}(a) and (c)) a longer disturbance imparts less energy to proper modes and the wave train disperses more quickly, with a stronger decrease of its amplitude. On the other hand, a longer transverse disturbance (compare Figures~\ref{fig_comp_kink_sausage}(b) and (d)) imparts less energy to the first overtone (and so the trailing oscillations are missing in panel (d)) and the fundamental mode receives more energy in the long-wavelength regime, for which $\cg$ has a small spread. This then results in less dispersion of the wave train. It is also worth noting that the first fast kink mode overtone leaves its imprint in Figures~\ref{fig_comp_kink_sausage}(b) and (f) as small amplitude, short period oscillations that arrive at the detection point after the main pulse. Finally, the small amplitude oscillations in panels (a) and (c) before and after the wave train passage are numerical artifacts.

\subsection{Wave Leakage}\label{sect_leakage}

The improper modes are closely related to leaky quasi-modes studied by, e.g., \citet{cally1986,cally2003}; see also the discussion by \citet{andries2007}. The interference of improper modes will create a wave front propagating from the tube in the radial direction. The leaky quasi-modes describe the so-called intermediate asymptotic of this wave front \citep[see][ for a discussion of the role of leaky quasi-modes in the case of kink waves]{ruderman2006}.

\begin{figure*}[ht!]
  \centerline{
    \scriptsize{(a)}\hspace{-10ex}
    \includegraphics[width=0.35\textwidth,angle=-90]{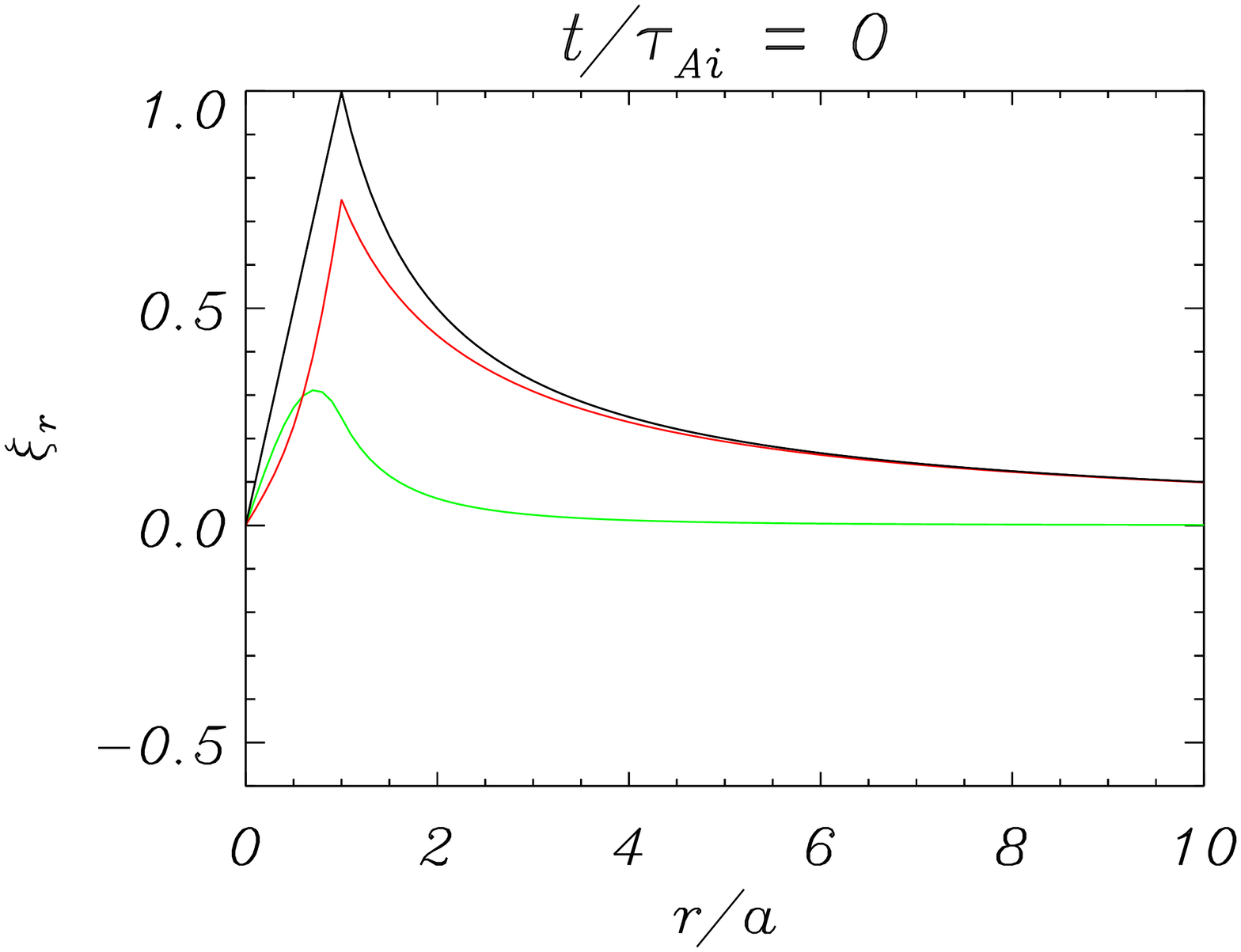} \\
    \scriptsize{(b)}\hspace{-10ex}
    \includegraphics[width=0.35\textwidth,angle=-90]{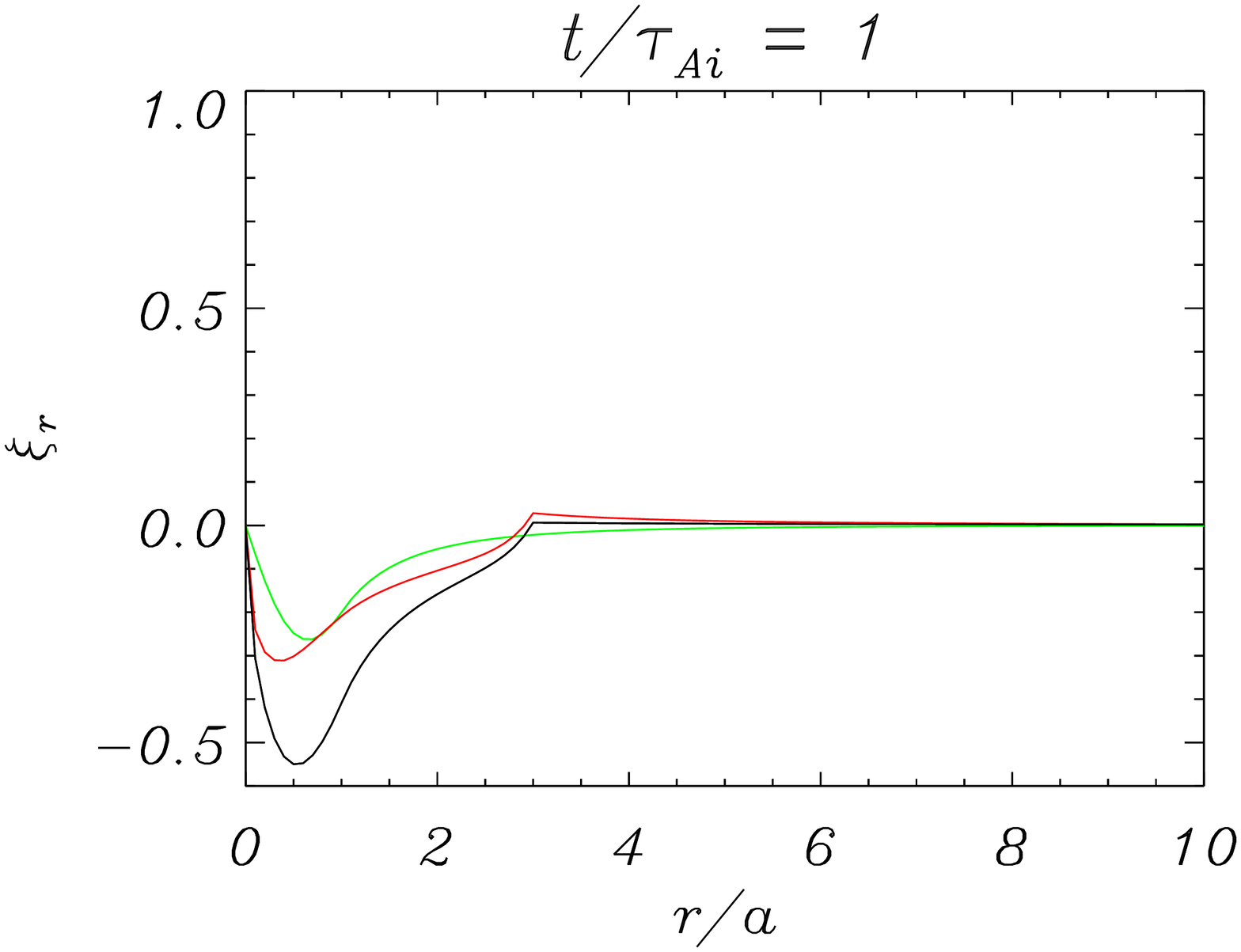} \\
  }
  \centerline{
    \scriptsize{(c)}\hspace{-10ex}
    \includegraphics[width=0.35\textwidth,angle=-90]{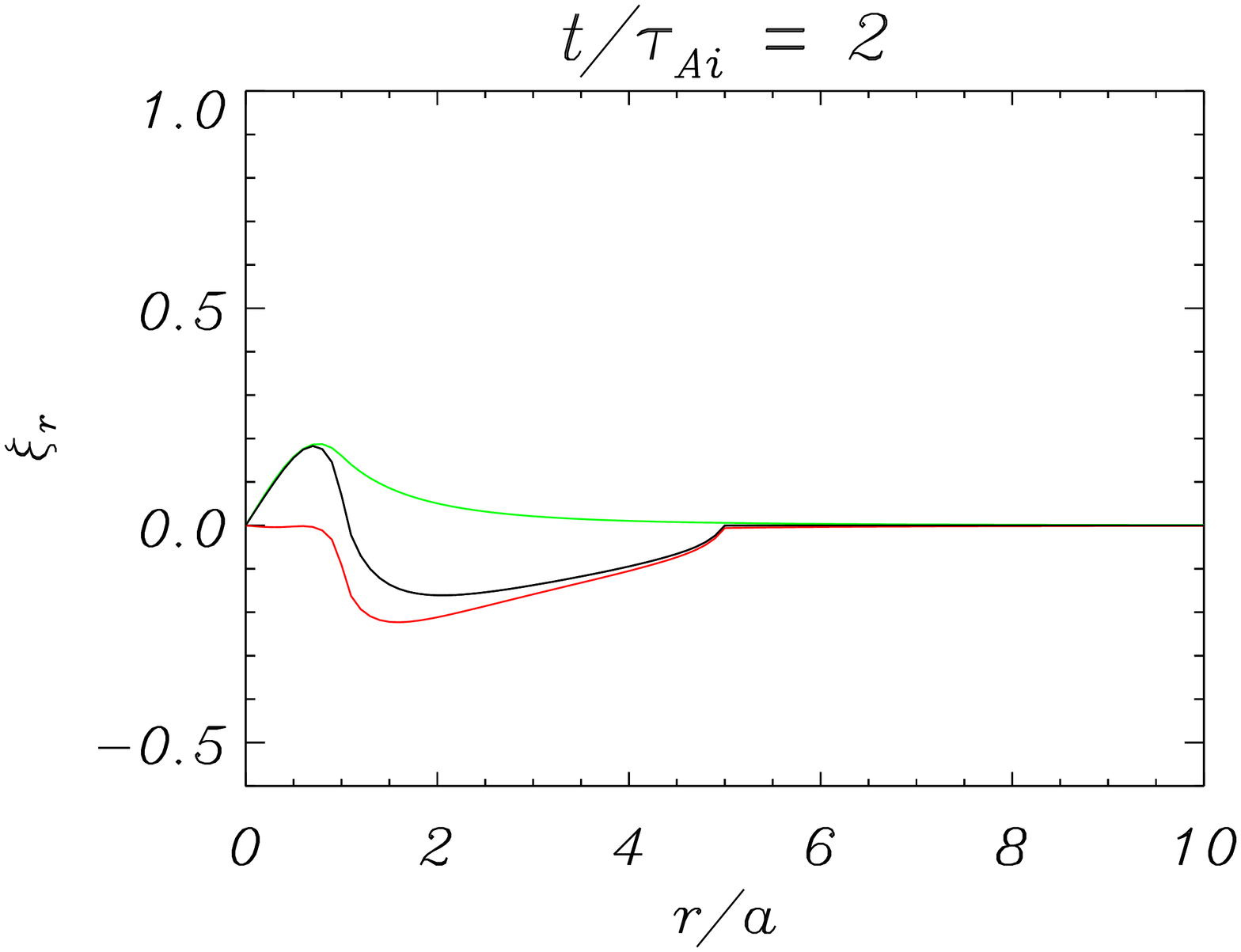} \\
    \scriptsize{(d)}\hspace{-10ex}
    \includegraphics[width=0.35\textwidth,angle=-90]{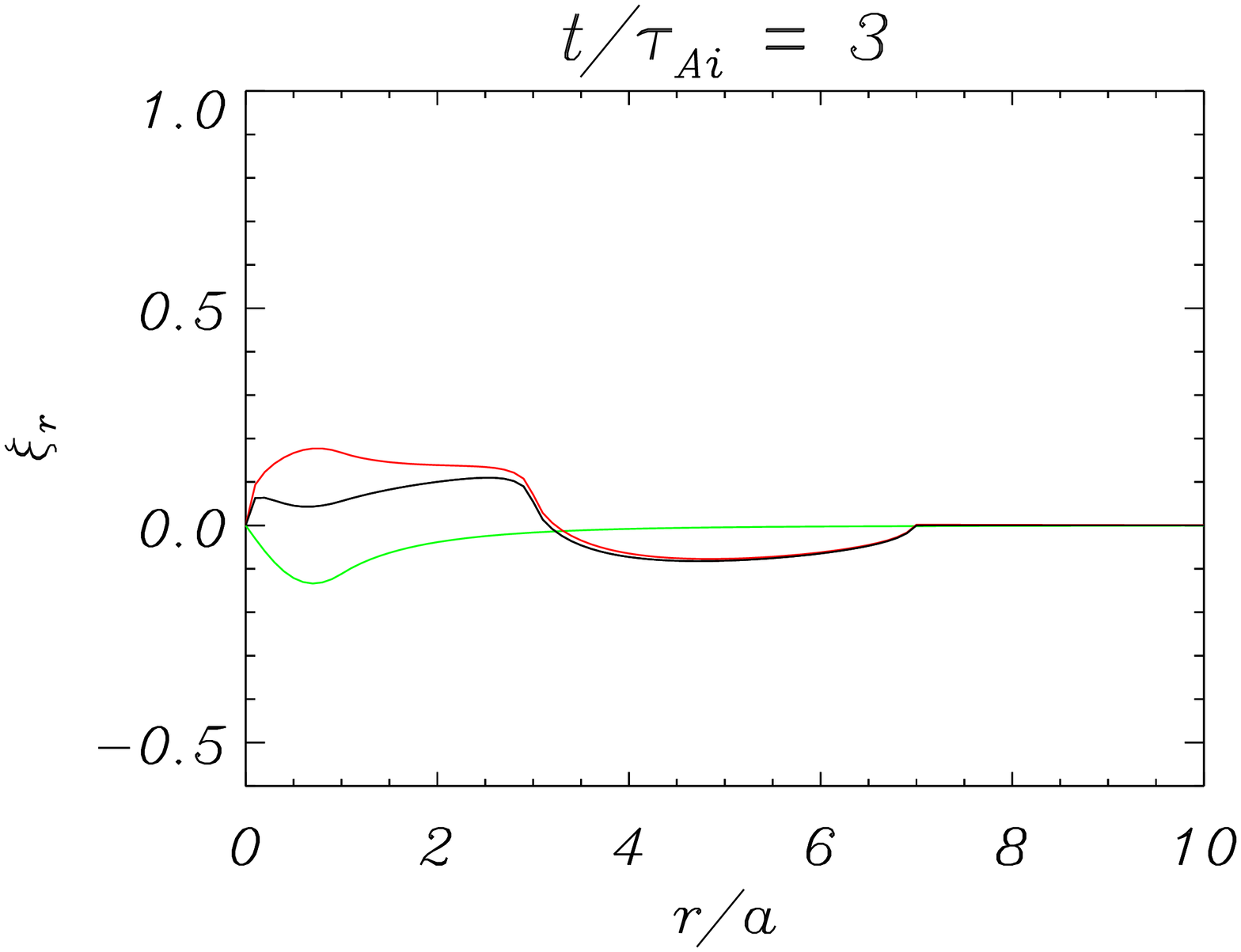} \\
  }
  \centerline{
    \scriptsize{(e)}\hspace{-10ex}
    \includegraphics[width=0.35\textwidth,angle=-90]{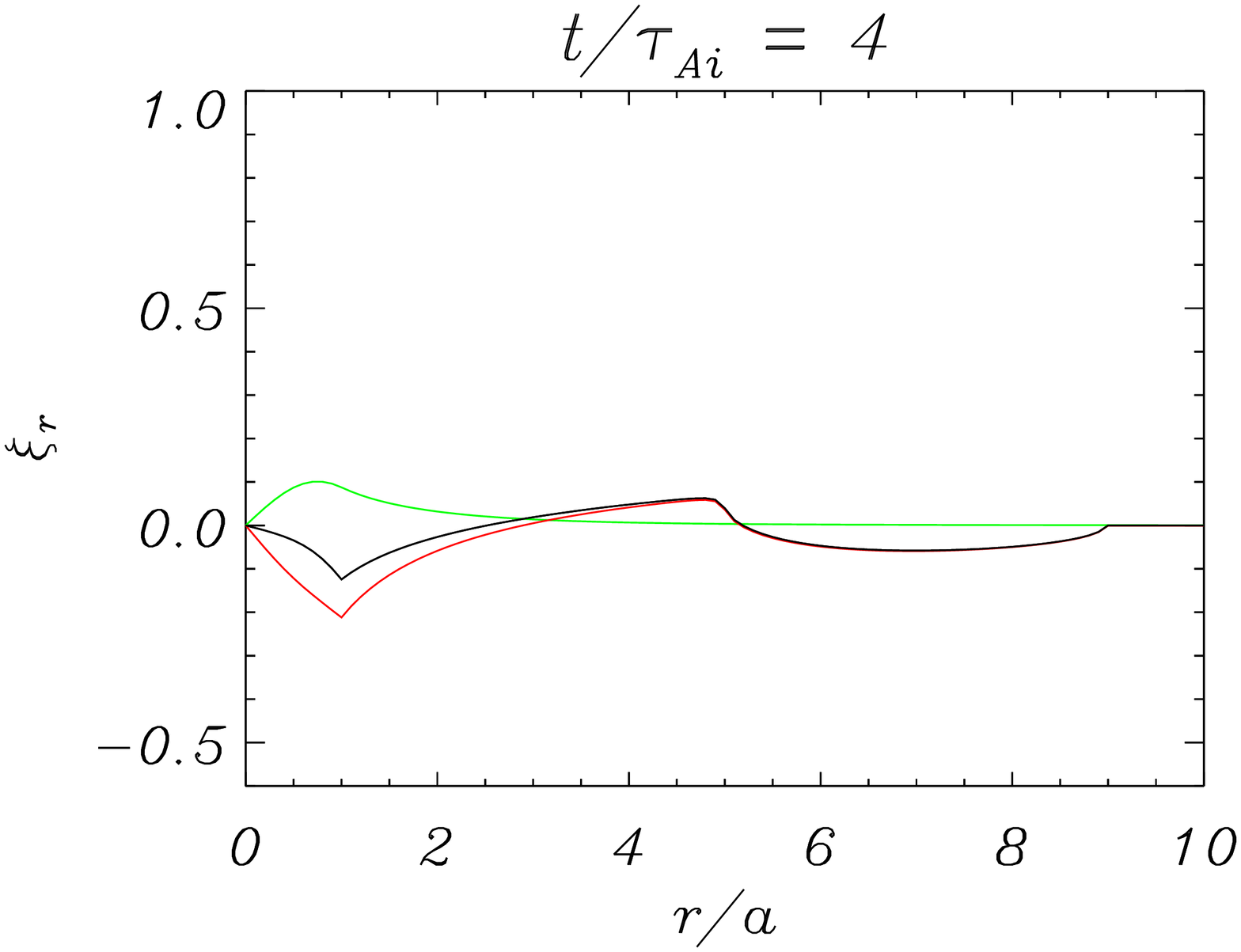} \\
    \scriptsize{(f)}\hspace{-10ex}
    \includegraphics[width=0.35\textwidth,angle=-90]{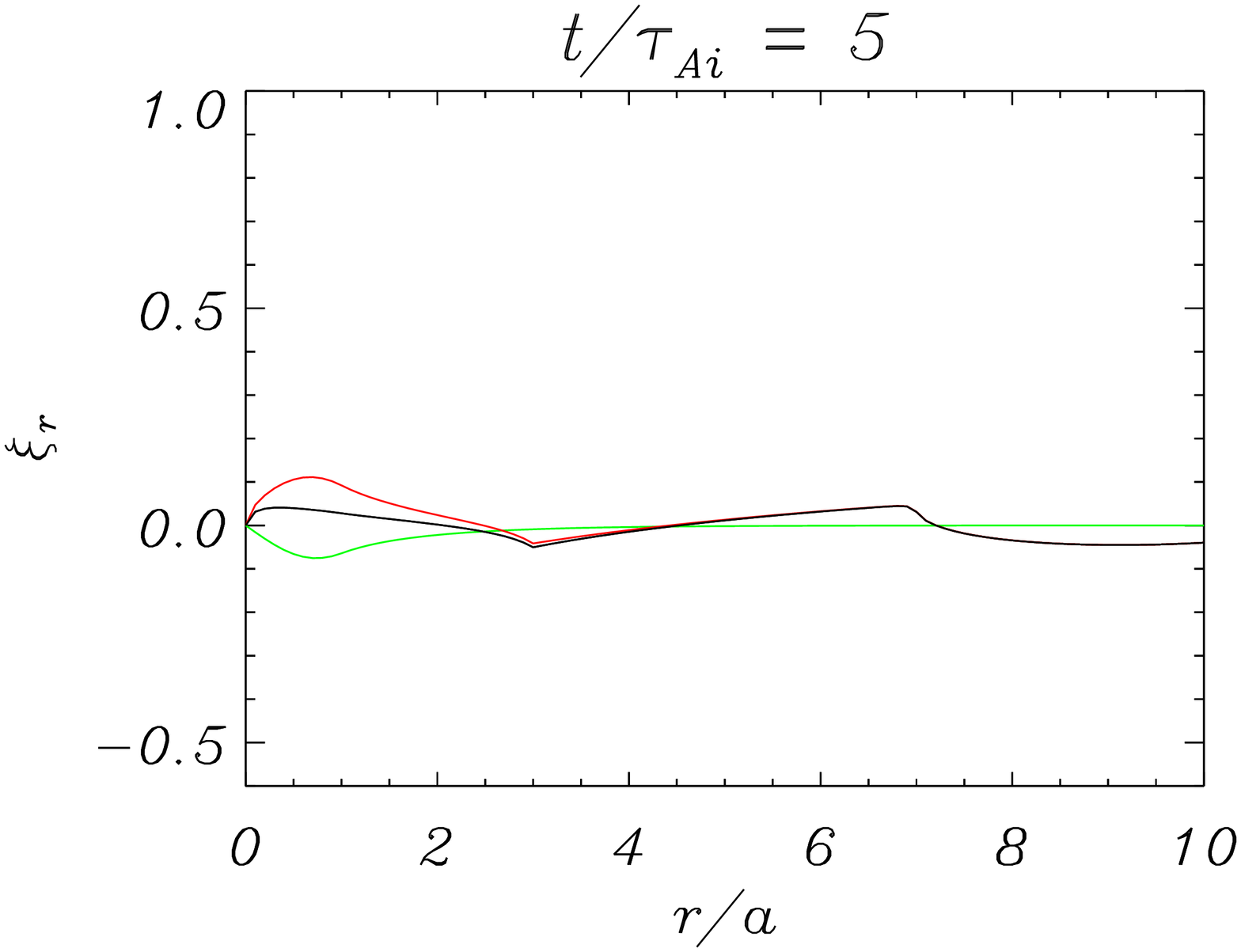}
  }
  \caption{Radial displacement of the plasma as a function of radial distance at $z=0$ and for different times, shown above each frame. Black lines correspond to the displacement $\xi_r$, while green and red lines correspond to the contributions of proper and improper modes, respectively. The density ratio is $\rhoi/\rhoe=4$ and the length of the initial disturbance is of the order of $4a$.}
  \label{fig_leakage}
\end{figure*}

Wave radiation after the initial excitation can be well appreciated by plotting snapshots of the dependence of $\xi_r$ with respect to $r$ for several times (Figure~\ref{fig_leakage}). Panel (a) shows the imposed radial profile of the radial displacement at $t=0$, given by Equation~(\ref{init_psi}), with improper modes (red curve) having a more important contribution than proper modes (green curve). The subsequent temporal evolution of $\xi_r$ shows oscillations inside the magnetic cylinder ($0\leq r/a\leq 1$) and in its vicinity, together with a wave pulse propagating outwards in the radial direction, that is specially visible in panel (d) as a depression in $\xi_r$ in the range $3\lesssim r/a\lesssim7$. This propagating feature is clearly caused by improper modes because the amplitude of proper modes at its position is negligible.

The wave emission process has been presented here by doing the cuts of Figure~\ref{fig_leakage} at the center of the initial perturbation ($z=0$), but as far as the amplitude of improper modes is non zero, it also takes place at other positions along the magnetic tube when they are affected by the propagating disturbance.

\section{DISCUSSION}\label{sect_conclusions}

In this paper we have used a method based on Fourier integrals to solve the initial value problem of impulsively excited axisymmetric (i.e., sausage) perturbations in a uniform, dense, cylindrical magnetic tube. The linear and $\beta=0$ approximations have been assumed. This method gives analytical expressions for the perturbations (plasma displacement, density, velocity, and magnetic field) that require the numerical computation of the two integrals in Equations~(\ref{expand_eigen-main}) and (\ref{Four_trans_inv-main}). The accuracy of the results depends exclusively on the precision of the numerical computation of these integrals. An advantage of this method is that one can select which perturbations will be computed and at which spatial position(s) and time(s).

The impulsive generation of fast sausage wave trains in a dense magnetic tube has also been studied before by numerically integrating the linearized MHD equations. Most of these works use a Cartesian slab to represent the magnetic tube \citep{murawski1993b,murawski1993d,nakariakov2004,nakariakov2005,jelinek2010,nakariakov2012,meszarosova2014}. In the solar context, only \citet{selwa2004} have considered this problem in cylindrical geometry, although they did not take into account the dispersive behavior of the propagating wave packet.

One of the key features of the present paper is the importance of the group velocity in the dispersion of a concentrated perturbation. Concerning fast sausage wave trains, this was already acknowledged by \citet{roberts1984} and later by other authors \citep[e.g.,][]{nakariakov1995,nakariakov2004}. In addition to this, we have emphasized the role of the length of the initial perturbation and the internal to external density ratio in the wave packet dispersion (Figures~\ref{fig_longer_t} and \ref{fig_comp_kink_sausage}). We have found that longer initial perturbations have much lower amplitudes because the initial excitation imparts less energy to proper modes. In addition, larger values of the density ratio lead to a wider wave packet that nevertheless has a larger amplitude because most of the oscillatory energy is concentrated in its trailing part. Some of these results have also been found in a non-uniform dense slab by \citet{nakariakov2005}

We are now concerned with the observational implications of our work. One of the main conclusions of Paper~I is that a spatially concentrated, transverse disturbance of a magnetic tube gives rise to non-negligible density perturbations that take their maximum value at the tube boundary. The reason for these density perturbations is that fast kink waves are compressive for longitudinal wavelengths of the order of the tube radius. The axisymmetric wave trains studied in the present work also generate density perturbations caused by the compressive character of fast sausage waves. To determine whether the wave trains propagating along coronal loops studied by \citet{williams2001,williams2002,katsiyannis2003} \citep[and similar events detected in solar radio bursts, see][]{meszarosova2009,meszarosova2011,karlicky2013} are caused by fast sausage or kink excitations it is necessary to carry out a forward modelling of the magnetohydrodynamic perturbations to determine how they would appear when observed with available instruments. \citet{cooper2003a,cooper2003b} assumed that the observed emission intensity in optically thin coronal lines is proportional to the line integral along the line-of-sight (LOS) of the electron number density squared. They considered fast sausage and kink waves with fixed wavelength and different orientations of the magnetic tube axis with respect to the LOS. They concluded that the two wave modes generate similar amplitudes of integrated density variations. Nevertheless, the integrated intensity alone does not govern the line intensity and formation of emission lines. \citet{demoortel2008} took into account that it is actually governed by a complex, non-linear interaction between the density squared, the ionisation balance and emissivity, and the instrument response function. The important point here is that the ionisation equilibrium timescale of \ion{Fe}{10} and \ion{Fe}{12} ions in a coronal environment is of the order of a few to tens of seconds. The effect of the ionisation balance can be ignored only if the wave period is much longer than this timescale. Otherwise, observed intensity perturbations do not necessarily behave like the theoretical temperature and density. Therefore, to make a theoretical interpretation of the $\sim 6$~s oscillations described by \citet{williams2001,williams2002,katsiyannis2003} a careful forward modelling of impulsively excited fast sausage and kink perturbations is required. \citet{antolin2013} \citep[see also][]{reznikova2014} studied the influence of LOS angle and spatial, temporal, and spectral resolution of observing instruments on fast sausage wave observables. Under the assumption of equilibrium ionisation, these authors concluded that the intensity variation can be very low ($\lesssim 4$\% for \ion{Fe}{9} 171 \AA) or significant ($\lesssim 35$\% for \ion{Fe}{12} 193 \AA). Most of the intensity variation, however, disappears when considering the radiative relaxation times of the ions. Another important conclusion of this work is that the intensity modulation has strong dependence on the LOS angle and on the spatial and temporal resolutions. Our future aim is to repeat the calculations of \citet{antolin2013} for impulsive initial excitations of the fast sausage and kink types.


\section*{Acknowledgements}

J.T. acknowledges support from the Spanish Ministerio de Educaci\'on y Ciencia through a Ram\'on y Cajal grant. R.O. and J.T. also acknowledge the funding provided under the project AYA2011-22846 by the Spanish MICINN and FEDER Funds. R.O. is indebted to D. W. Fanning for making available the Coyote Library of IDL programs (http://www.idlcoyote.com/) and to W. Rosenheinrich for his table of Bessel functions integrals (http://www.fh-jena.de/$\sim$rsh/Forschung/Stoer/besint.pdf). M.S.R. also acknowledges the support by the STFC grant.

\appendix

\section{FREQUENCY OF PROPER MODES FOR LARGE \lowercase{$k$}}\label{app_large_k}

Our purpose here is to study the solutions to the dispersion relation
\begin{equation}
\frac{J_1(\ki a)}{\ki a J_0(\ki a)} + \frac{K_1(\kappae a)}{\kappae a K_0(\kappae a)} = 0
\label{appA:eq:1}
\end{equation}
in the limit of very large longitudinal wavenumber, i.e., for $ka\gg 1$. We make use of the asymptotic formulae for large arguments \citep{abramowitz1964}
\begin{equation}
K_0(x) \sim K_1(x) \sim \sqrt{\frac\pi{2x}}e^{-x} .
\label{appA:eq:2}
\end{equation}
Figure~\ref{fig_dr_over} shows that, for $ka \gg 1$, $\omega/k$ is not close to $\vae$ and so $\kappae a\gg 1$. Then, the second term in the dispersion relation is
\begin{equation}
\frac{K_1(\kappae a)}{\kappae aK_0(\kappae a)} \approx \frac1{\kappae a} \ll 1.
\label{appA:eq:3}
\end{equation}
Again using Figure~\ref{fig_dr_over} we see that, for $ka \gg 1$, $\omega/k$ is close to $\vai$ and so $\ki$ is not large. Since the second term in Equation~(\ref{appA:eq:1}) is small, the first one also has to be small. This is only possible when $\ki a$ is close to one of the zeros of the function $J_1(x)$, which are $j_{1j}$\/, $j = 1,2,\dots$ As a result, we have
\begin{equation}
\omega^2 = k^2 \vai^2\left(1 + \frac{\rhoi j_{1j}^2}{\rhoe k^2 a^2}\right).
\label{appA:eq:5}
\end{equation}
From this expression, the phase speed and group velocity for large longitudinal wavenumber are
\begin{equation}
\cph\approx\vai\left(1 + \frac{\rhoi j_{1j}^2}{2\rhoe k^2 a^2}\right), \quad \cg\approx\vai\left(1 - \frac{\rhoi j_{1j}^2}{2\rhoe k^2 a^2}\right).
\end{equation}
We conclude that as $ka$ tends to infinity, $\cph$ tends to $\vai$ from above and $\cg$ tends to $\vai$ from below.

\section{CALCULATION OF \lowercase{$q(\omega)$}}\label{app_q_of_omega}

Using Equation~(\ref{xir_of_r_im}) we write the expression for the scalar product of two improper eigenfunctions as 
\begin{align}
\langle\hat{\xi}_\omega,\hat{\xi}_{\omega'}\rangle &= 
   \frac{\rho_i a^4 k_e^2{k'_e}^2}{k_i k'_i} \int_0^a J_1(k_i r) J_1(k'_i r)\,r\,dr \nonumber \\
& +\rho_e a^4 k_e k'_e \int_a^\infty F_1(k_e r)F_1(k'_e r)\,r\,dr ,
\label{eq:6}
\end{align}
where
\begin{equation}
F_1(x) = C_J J_1(x) + C_Y Y_1(x) .
\label{eq:7}
\end{equation}
Now we use the formulae
\begin{align}
& J_1(z) = \sqrt{\frac2{\pi z}}\cos\left(z - \mbox{$\frac34$}\pi\right)
   \left[1 + j_1(z)\right] , \nonumber \\
& Y_1(z) = \sqrt{\frac2{\pi z}}\sin\left(z - \mbox{$\frac34$}\pi\right)
   \left[1 + y_1(z)\right] ,
\label{eq:8}
\end{align}
where $j_1(z)$ and $y_1(z)$ decay as $z^{-1}$ when $z \to \infty$\/. Using these formulae, after some algebra, we rewrite Equation~(\ref{eq:6}) as
\begin{equation}
\langle\hat{\xi}_\omega,\hat{\xi}_{\omega'}\rangle =
   \Gamma(\omega,\omega') + \Upsilon(\omega,\omega') ,
\label{eq:9}
\end{equation}
where 
\begin{equation}
\Gamma(\omega,\omega') = \Gamma_1(\omega,\omega') + \Gamma_2(\omega,\omega') + \Gamma_3(\omega,\omega'),
\label{eq:10}
\end{equation}
\begin{equation}
\Gamma_1(\omega,\omega') = 
   \frac{\rho_i a^4 k_e^2{k'_e}^2}{k_i k'_i} \int_0^a J_1(k_i r) J_1(k'_i r)\,r\,dr ,
\label{eq:11}
\end{equation}
\begin{align}  
\Gamma_2(\omega,\omega') &= \frac{\rho_e a^4\sqrt{k_e k'_e}}\pi \nonumber \\
& \times \int_a^\infty
   \left\{\left[C_J C'_J j_1(k'_e r) + C_J C'_Y y_1(k'_e r)\right]\right. \nonumber \\
& \hspace{10ex} \times \cos\left(k_e r - \mbox{$\frac34$}\pi\right) \nonumber \\
& \hspace{8ex} + \left[C_Y C'_J j_1(k'_e r) + C_Y C'_Y y_1(k'_e r)\right] \nonumber \\
& \hspace{10ex} \times \sin\left(k_e r - \mbox{$\frac34$}\pi\right) \nonumber \\
& \hspace{8ex} + \left[C_J C'_J j_1(k_e r) + C_Y C'_J y_1(k_e r)\right] \nonumber \\
& \hspace{10ex} \times \cos\left(k'_e r - \mbox{$\frac34$}\pi\right) \nonumber \\
& \hspace{8ex} + \left[C_J C'_Y j_1(k_e r) + C_Y C'_Y y_1(k_e r)\right] \nonumber \\
& \hspace{10ex} \times \left. \sin\left(k'_e r - \mbox{$\frac34$}\pi\right)\right\}\,dr ,
\label{eq:12} 
\end{align} 
%
\begin{align}   
\Gamma_3(\omega,\omega') = \frac{\rho_e a^2}{\pi\sqrt{k_e k'_e}} 
   \int_a^\infty & \left[C_J C'_J j_1(k_e r) j_1(k'_e r) \right. \nonumber \\
   &+  C_J C'_Y j_1(k_e r) y_1(k'_e r) \nonumber \\
& + C_Y C'_J y_1(k_e r) j_1(k'_e r) \nonumber \\
& + \left. C_Y C'_Y y_1(k_e r) y_1(k'_e r)\right] dr ,
\label{eq:13}
\end{align}
\begin{align}
\Upsilon(\omega,\omega') &= \frac{\rho_e a^4}\pi\sqrt{k_e k'_e} \nonumber \\
& \times \int_a^\infty\left\{(C_J C'_J + C_Y C'_Y)
   \cos\left[r(k_e - k'_e)\right]\right.  \nonumber \\
& \hspace{8ex} - (C_J C'_Y + C_Y C'_J)\cos\left[r(k_e + k'_e)\right] \nonumber \\
& \hspace{8ex} - (C_J C'_Y - C_Y C'_J)\sin\left[r(k_e - k'_e)\right] \nonumber \\
& \hspace{8ex} + \left. (C_J C'_J - C_Y C'_Y)\sin\left[r(k_e + k'_e)\right]\right\}dr .
\label{eq:13b}
\end{align}
As it is explained in Paper~I, $\Gamma(\omega,\omega')$ is a regular function of $\omega$ and $\omega'$\/. Repeating the same calculation as in Paper~I we obtain that
\begin{equation}
\langle\hat{\xi}_\omega,\hat{\xi}_{\omega'}\rangle = 
    Q(\omega,\omega') + \frac{\rho_e a^4 \vae^2 k_e^2}\omega 
    W(\omega,\omega')\delta(\omega - \omega'),
\label{eq:14}
\end{equation}
where $Q(\omega,\omega')$ is a regular function and
\begin{equation}
W(\omega,\omega') = (C_J C'_J + C_Y C'_Y)\cos\left[a(k_e - k'_e)\right] .
\label{eq:15}
\end{equation}
Using the same arguments as in Paper~I we prove that $Q(\omega,\omega') \equiv 0$ and $W(\omega,\omega')\delta(\omega - \omega') = W(\omega,\omega) \delta(\omega - \omega')$. Then we obtain
\begin{equation}
q(\omega) = \frac{\rho_e a^4 \vae^2 k_e^2}\omega\left(C_J^2 + C_Y^2\right) .
\label{eq:16}
\end{equation}

\section{AMPLITUDES OF FAST KINK MODES FOR SPECIFIC INITIAL CONDITIONS}\label{app_kink}

In Paper~I the evolution of an initial transverse displacement of the magnetic tube was studied. The radial variation of the initial disturbance outside the magnetic cylinder was taken as $\exp(-(r-a)^2/l^2)$, with $l$ a prescribed length. In the present paper this external radial variation is $a/r$. We now derive expressions for the amplitudes $A^\pm_j(k)$ and $A^\pm_\omega(k)$ for a transverse perturbation whose initial radial distribution is $\psi(r)=a/r$ for $r\geq a$. Note that for this type of perturbation we still use $\psi(r)=1$ for $r\leq a$.

We start with the proper mode amplitudes $A_j^\pm(k)$, that are given by Equation~(\ref{A_pm-main}) with

\begin{align}\label{nj_kink}
\mathcal{N}_j(k) &=a\xi_0\sqrt{\pi}\Delta\omega_j\exp(-\Delta^2k^2/4)J_1(\ki a) \nonumber \\
& \times \bigg\{\frac{\rhoi}{\ki^2}K_1(\kappae a) -\frac{\rhoe}{\kappae^2}\bigg[-\frac{\pi}{2}+\kappae aK_0(\kappae a) \nonumber \\
& \hspace{22.5ex} +\Psi_K(\kappae a)\bigg]\bigg\},
\end{align}

\noindent where

\begin{equation}
\Psi_K(x)=\frac{\pi}{2}x\left[K_0(x){\bf L}_1(x)+K_1(x){\bf L}_0(x)\right].
\end{equation}

\noindent This function is introduced by \citet{rosenheinrich}, p.~6, and appears when obtaining the primitive of $K_1(x)/x$, see \citet{rosenheinrich}, p.~19. Furthermore, ${\bf L}_0(x)$ and ${\bf L}_1(x)$ are the modified Struve functions.

The denominator of Equation~(\ref{A_pm-main}) contains

\begin{align}\label{dj_kink}
\mathcal{D}_j(k) &=\frac{\rhoi K_1^2(\kappae a)}{2\ki^4} \big\{\ki^2a^2J_0^2(\ki a) \nonumber \\
& \hspace{15ex} +\left(\ki^2a^2-2\right)J_1^2(\ki a)\big\} \nonumber \\
& +\frac{\rhoe J_1^2(\ki a)}{2\kappae^4} \big\{-\kappae^2a^2K_0^2(\kappae a) \nonumber \\
& \hspace{14.5ex} +\left(\kappae^2a^2+2\right)K_1^2(\kappae a)\big\}.
\end{align}

\noindent To derive this result the primitive of $xK_1'^2(x)$ is required. We have used Equation~(\ref{derivJ1K1}) to express $K_1'(x)$ in terms of $K_0(x)$ and $K_1(x)$. Next, the primitive has been calculated with the help of expressions in \citet{rosenheinrich}, pp.~143 and 160, with $I_0(x)$ and $I_1(x)$ replaced by $K_0(x)$ and $K_1(x)$.

Finally, Equation~(\ref{A_pm_om-main}) gives the amplitudes $A^\pm_\omega(k)$, where the numerator is

\begin{align}\label{nomega_kink}
\mathcal{N}_\omega(k) &=a\xi_0\sqrt{\pi}\Delta\omega_j\exp(-\Delta^2k^2/4) \nonumber \\
& \times \bigg\{\frac{\rhoi\ke^2a^2}{\ki^2}J_1(\ki a) \nonumber \\
& \hspace{4.5ex} +\rhoe a^2\bigg[C_J\big(1-\ke aJ_0(\ke a)-\Phi(\ke a)\big) \nonumber \\
& \hspace{13ex} +C_Y\big(-\ke aY_0(\ke a)-\Phi_Y(\ke a)\big)\bigg]\bigg\},
\end{align}

\noindent where the primitive of $J_1(x)/x$ is given in \citet{rosenheinrich}, p.~19, in terms of the function

\begin{equation}
\Phi(x)=\frac{\pi}{2}x\left[J_1(x){\bf H}_0(x)-J_0(x){\bf H}_1(x)\right],
\end{equation}

\noindent with ${\bf H}_0(x)$ and ${\bf H}_1(x)$ the Struve functions. The primitive of $Y_1(x)/x$ is derived in the same way by substituting $J_0(x)$ and $J_1(x)$ by $Y_0(x)$ and $Y_1(x)$ and $\Phi(x)$ by $\Phi_Y(x)$. This function is defined as

\begin{equation}
\Phi_Y(x)=\frac{\pi}{2}x\left[Y_1(x){\bf H}_0(x)-Y_0(x){\bf H}_1(x)\right].
\end{equation}

The numerical evaluation of the Struve and modified Struve functions has been done with the software developed by \citet{macleod1996}.


\end{document}